\documentclass[10pt,nc,aps,notitlepage,superscriptaddress,twocolumn,longbibliography]{revtex4-2}

\usepackage{amsmath}
\usepackage{amssymb}
\usepackage{amsfonts}
\usepackage{bbm}
\usepackage{bm}
\usepackage{graphicx}
\usepackage{color}
\usepackage{dcolumn}
\usepackage{MnSymbol}
\usepackage{times}
\usepackage{braket}

\providecommand{\tr}{\operatorname{Tr}}
\providecommand{\op}[2]{\ket{#1}\bra{#2}}
\providecommand{\inp}[1]{\braket{#1}}

\usepackage{listings}
\newcommand*{\id}{\mathbbm{1}}
\newcommand*{\Tr}{\textrm{Tr}}

\usepackage{color}
\definecolor{dred}{rgb}{.8,0.2,.2}
\definecolor{ddred}{rgb}{.8,0.5,.5}
\definecolor{dblue}{rgb}{.2,0.2,.8}
\definecolor{dgreen}{rgb}{.2,0.5,.2}

\usepackage{amsthm,amsmath,amssymb}
\usepackage{mathrsfs}
\newcommand*{\physus}{Department of Physics and Guangdong Basic Research Center of Excellence for Quantum Science, Southern University of Science and Technology, Southern University of Science and Technology, Shenzhen 518055, China}
\newcommand*{\inssus}{Shenzhen Institute for Quantum Science and Engineering, Southern University of Science and Technology, Shenzhen 518055, China}

\usepackage[small,raggedright]{titlesec}
\titleformat{name=\section,numberless}{\normalsize\sffamily\bfseries}{}{0em}{}
\titleformat{name=\subsection,numberless}{\normalsize\sffamily\bfseries}{}{0em}{}
\titleformat*{\section}{\small\bfseries}
\titlespacing*{\section}{0pt}{10pt}{0pt}
\titlespacing*{\subsection}{0pt}{10pt}{0pt}

\usepackage[colorlinks=true,linkcolor=blue,citecolor=blue,urlcolor=blue]{hyperref}
\usepackage[capitalise]{cleveref}
\setcitestyle{numbers,round,comma,sort&compress}
\makeatletter

\def\NAT@nmfmt#1{\textit{#1}}
\makeatother
\hypersetup{citecolor=black}

\usepackage{etoolbox}

\makeatletter
\renewcommand{\bibinfo}[2]{%
  \ifstrequal{#1}{journal}{\textit{#2}}{#2}}
\makeatother

\begin{document}

\title{Experimental causal inference from measurements alone in the presence of spatial and temporal quantum correlations
}

\author{Hongfeng Liu}
\thanks{These authors contributed equally to this work.}
\affiliation{\physus}
%12331036@mail.sustech.edu.cn

\author{Xiangjing Liu}
\thanks{These authors contributed equally to this work.}
\affiliation{Nanyang Quantum Hub, School of Physical and Mathematical Sciences, Nanyang Technological University, Singapore }
\affiliation{Centre for Quantum Technologies, Nanyang Technological University, 637371, Singapore}

%xiangjing.liu@ntu.edu.sg

\author{Qian Chen}
\affiliation{Mathematics Research Center, School of Science and Engineering, The Chinese University of Hong Kong, Shenzhen, China}
\affiliation{QICI Quantum Information and Computation Initiative, School of Computing and Data Science, The University of Hong Kong, Pokfulam Road, Hong Kong}
\affiliation{Univ Lyon, Inria, ENS Lyon, UCBL, LIP, F-69342, Lyon Cedex 07, France}
%chenqian.phys@gmail.com

\author{Yixian Qiu}
\affiliation{Centre for Quantum Technologies, National University of Singapore, Singapore 117543, Singapore}
%yixian_qiu@u.nus.edu

\author{Vlatko Vedral}
\affiliation{Clarendon Laboratory, University of Oxford, Parks Road, Oxford OX1 3PU, United Kingdom}
%vlatko.vedral@physics.ox.ac.uk

\author{Xinfang Nie}
\email{ Corresponding author:  niexf@sustech.edu.cn}
\affiliation{\physus}
\affiliation{Quantum Science Center of Guangdong-Hong Kong-Macao Greater Bay Area, Shenzhen 518045, China}

\author{Oscar Dahlsten}
\email{Corresponding author: oscar.dahlsten@cityu.edu.hk}
\affiliation{Department of Physics, City University of Hong Kong, Tat Chee Avenue, Kowloon, Hong Kong SAR, China}
%\affiliation{\physus}
\affiliation{\inssus}
\affiliation{Institute of Nanoscience and Applications, Southern University of Science and Technology, Shenzhen 518055, China}

\author{Dawei Lu}
\email{Corresponding author: ludw@sustech.edu.cn}
\affiliation{\physus}
\affiliation{Quantum Science Center of Guangdong-Hong Kong-Macao Greater Bay Area, Shenzhen 518045, China}
\affiliation{\inssus}

\date{\today}

\begin{abstract} 
\noindent {\bf Abstract}\\
The causal structure of a set of events specifies which events influence which others. Determining this structure, known as causal inference, is a fundamental task in science. Reset-style intervention is commonly used to determine causal structure in classical scenarios. However, in the more fundamental quantum theory, there is evidence that projective measurements alone can suffice in some cases. We here extend those cases to two systems at two times, more general channels, and from fine-grained to less invasive coarse-grained projective measurements. The measurements are implemented by scattering circuits in a nuclear magnetic resonance platform. The measurement data is analyzed via the pseudo-density matrix (PDM) formalism for measurements on systems at several times, in order to determine compatibility with given causal structures. We thereby successfully demonstrate causal inference from coarse-grained measurements alone, in a scenario with several qubits and two times.
\end{abstract}

\maketitle

%\tableofcontents

\section*{Introduction}
\begin{figure*}[t]
    \includegraphics[width=0.8\linewidth]{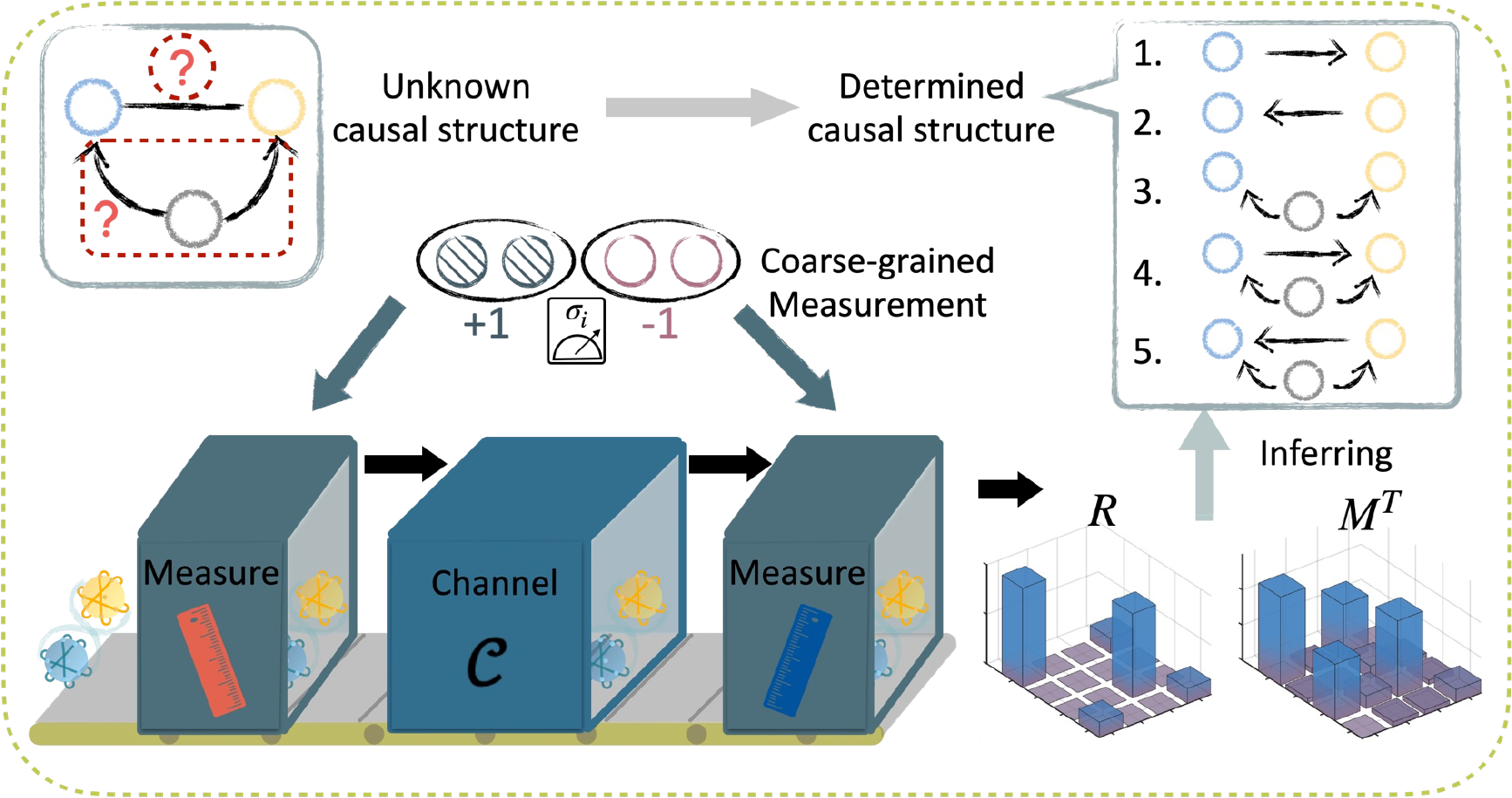}
    \caption{{\bf Overview of experiment and theoretical analysis.} It is unknown to the experimenter which quantum system is influencing which, and whether there are correlations in the initial state, termed a common cause. For example, structure 4 corresponds to correlations in the initial state followed by a channel whereby the blue system influences the yellow system. Data is taken from binary coarse-grained measurements on the initial state and after the channel. The data is turned into a pseudo density matrix $R$ and associated Choi matrix $M^{T_B}$, from which the causal structure is inferred.  }
\label{fig:Fig1} 
\end{figure*}

Determining the causal structure of a set of variables, i.e.\  which variables influence others~\cite{reichenbach1956direction, pearl2009causality}, often termed causal inference, is a fundamental task in science. That task has practical importance, for example, in medical trials~\cite{balke1997bounds, prosperi2020causal}, and in learning models of data that generalise well beyond the training data~\cite{peters2017elements,feder2022causal}. Reichenbach's Common Cause Principle identifies causal structures that may underlie probabilistic correlations. One version of the principle states that when there is a probabilistic correlation between two events, $A$ and $B$, then either (i) $A$ is a cause of $B$ (i.e.\ $A$ influences $B$), (ii) $B$ is a cause of $A$, or (iii) there is a third {\em common cause}, $C$, influencing both $A$ and $B$~\cite{reichenbach1956direction}. The three possibilities can also be mixed, leading to two more possibilities. The resulting five causal structures are depicted in Fig.~\ref{fig:Fig1}. Causal inference in this context then means determining which of those causal structures  is {\em compatible with the data} in that the structure could have generated the data~\cite{pearl2009causality}. In general, several structures can be compatible with the data. To gain additional data for causal inference, it is common in dealing with classical data to {\em intervene} in the system~\cite{pearl2009causality,balke1997bounds} to see how the probability distribution is altered. Partial inference from observations alone, without intervention, is an active area of research also for classical data, see e.g.\ Refs.~\cite{prosperi2020causal,pearl2009causality,angrist1996identification,peters2017elements, glymour2019review}.

Causal structure inference is also being investigated at the more fundamental quantum level~\cite{leggett1985quantum, oreshkov2012quantum,brukner2014quantum,fitzsimons2015quantum,barrett2021cyclic,hardy2005probability,costa2016quantum,allen2017quantum,PhysRevA.79.052110,Longxy2022,PRXQuantum.4.020334,PhysRevX.11.021043,song2023causal,wolfe2020quantifying,liu2024certifying, leifer2013towards, giarmatzi2019quantum}.  The causal structures of Reichenbach's principle were associated with quantum circuits~\cite{allen2017quantum}. Causal inference via interventions, in particular resetting the state of the quantum system, have been considered~\cite{ried2015quantum,bai2022quantum,chiribella2019quantum,maclean2017quantum,chaves2018quantum,agresti2022experimental, gachechiladze2020quantifying,nery2018quantum,agresti2020experimental}. It has been realized that, at least in some cases, such resets are not necessary to determine the causal structure and that, instead, observations in the form of projective measurements can suffice~\cite{ried2015quantum, liu2023inferring,fitzsimons2015quantum}. A quantum signature for temporal correlation derived from projective measurement data can rule out the common cause mechanism~\cite{fitzsimons2015quantum}. A theoretical scheme to distinguish certain causal structures (special cases of structures 1 and 3 in Fig.\ref{fig:Fig1}) from projective measurements was proposed and demonstrated in a photonic platform~\cite{ried2015quantum}. 
%for the case of a common cause, represented by correlations in the initial state, and a unitary channel.
Further, techniques to falsify if a given quantum causal model is compatible with some observed correlations were derived ~\cite{elie2021quantum,chaves2015information}. There are thus several instances in the quantum case where projective measurements are sufficient for causal inference, in contrast to the classical case where interventions resetting the state are generally needed. Of particular relevance here is recent theoretical results expanding the set of cases for which projective measurements suffice to two qubits at two times with general channels, and to coarse-grained projective measurements, which are less invasive than fine-grained~\cite{liu2023inferring,liu2023quantum}. A related experiment used randomized measurements to certify the presence of temporal correlations~\cite{liu2024certifying}; that approach detects whether such correlations exist but does not resolve the direction of causal influence, which is one of the goals of the present work.

We here adapt those protocols ~\cite{liu2023inferring,liu2023quantum} to provide the experimental demonstration of a quantum causal inference significantly expanding the set of causal structures that can be discriminated from measurements alone. The protocol is highly non-invasive in that the protocol uses only coarse-grained projective measurements. As shown in Fig.~\ref{fig:Fig1}, the observer gains data from joint measurements on two quantum systems (qubits) at two times. The joint measurements are coarse-grained projective measurements, implemented by unitary interaction with a probe qubit that later undergoes a fine-grained projective measurement. These coarse-grained measurements are implemented by a method termed scattering circuits~\cite{miquel2002interpretation,PhysRevA.105.L030402,Niexf2020otoc} in an NMR platform~\cite{fan2023experimental,lin2022experimental,Xin2021}. 
The protocol allows the observer to use the resulting data to infer the causal structure of the process generating the data or, more precisely, compatibility with a given causal structure or structures. For convenient analysis, the measurement data is turned into a pseudo-density matrix (PDM), a space-time state which extends the familiar density matrix for multiple systems to include multiple times~\cite{fitzsimons2015quantum,liu2023quantum}. We show, essentially, that two types of features of the relevant reduced state of the PDM are sufficient to determine compatibility with the causal structures of Fig.~\ref{fig:Fig1}: (i) negativity (ii) time-asymmetry. 

We successfully demonstrate the combination of those experimental and theoretical techniques to determine compatibility with the five causal structures of Fig.~\ref{fig:Fig1}, with subtleties including that we cannot distinguish between structures 4 and 5 and that we cannot determine the causal structure in a case of a fully decohering channel and a diagonal initial state. Our experimental demonstration significantly expands the set of cases for which projective, even the less invasive coarse-grained projective, measurements suffice for causal inference in the quantum case. We expand on the results in the Supplemental Material~\cite{supp}.

\section*{RESULTS}
\subsection*{Quantum causal inference task of five possible structures}
The observer has data from measurements on two quantum systems $A$ and $B$. The observer wants to know the causal structure generating the data. In line with Reichenbach's principle, we allow for five causal structures that are to be distinguished (see FIG.~\ref{fig:Fig1}). These structures are distinguished by the direction of any causal influence between $A$ (blue circle) and $B$ (yellow circle), and by whether there are initial correlations or not. (Scenarios with causal influence in both directions (loops), such as global unitaries on $A$ and $B$ are excluded, such that there is a well-defined causal direction~\cite{reichenbach1956direction,pearl2009causality}.)

\begin{figure}
    \includegraphics[width=1\linewidth]{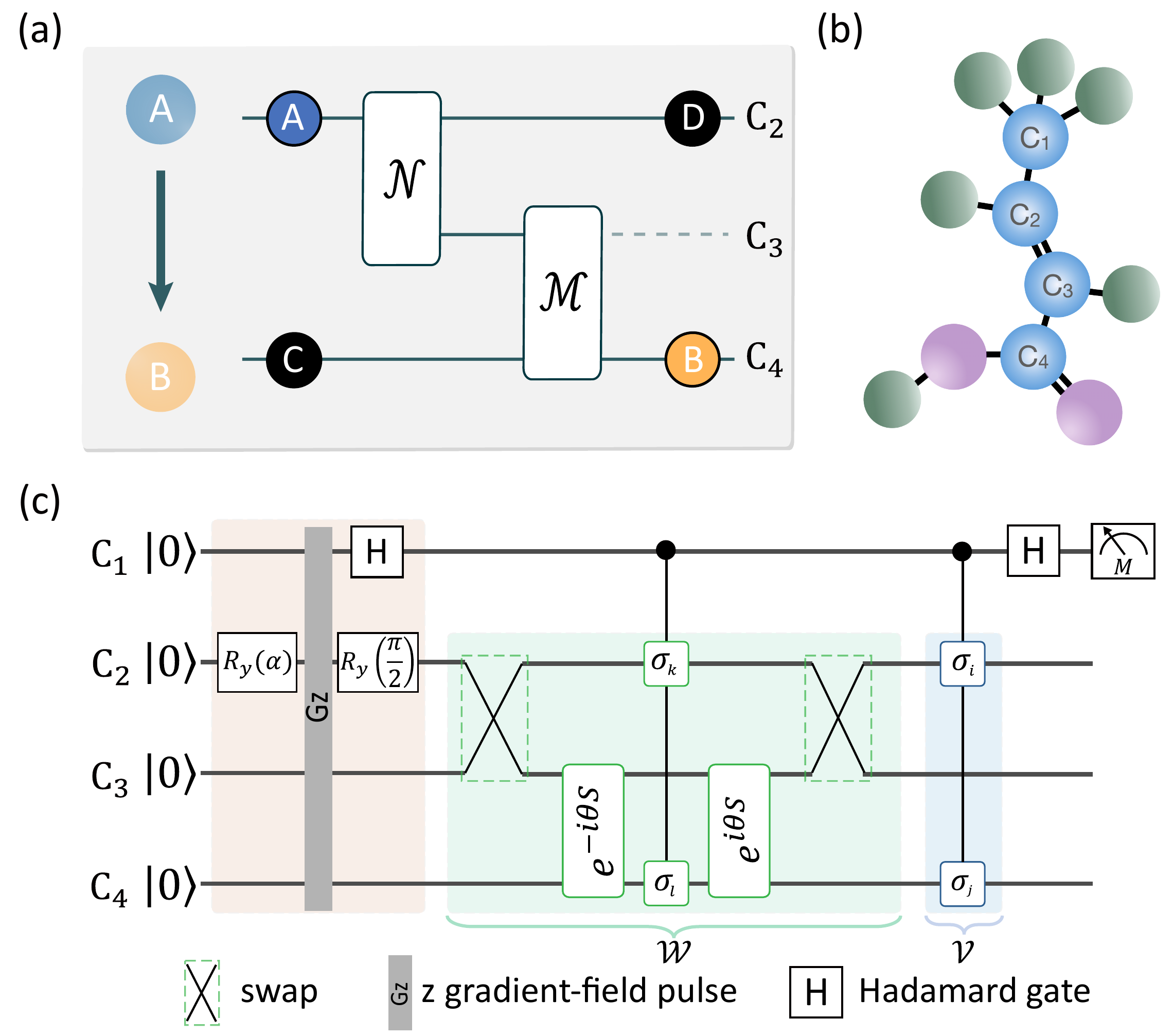}
    \caption{{\bf Experimental setup.} (a) Theoretical framework. The left panel shows the abstract causal structure $A\to B$ to be inferred, and the right panel shows the corresponding circuit model. The upper line (C$_2$) carries the event $A$ at time 1 and $D$ at time 2, while the lower line (C$_4$) carries $C$ at time 1 and $B$ at time 2. The two systems at different times are linked through consecutive channels $\mathcal{N}$ and $\mathcal{M}$ acting on an intermediate ancillary qubit (C$_3$). (b) Molecular structure of trans-crotonic acid, whose four $^{13}\text{C}$ nuclear spins encode the four qubits used in the NMR experiment. (c) Experimental quantum circuit implementing (a) on the four-qubit NMR processor. C$_1$ acts as the probe qubit of the scattering circuit (used to extract expectation values non-invasively), C$_2$ and C$_4$ are the two system qubits (corresponding to the upper and lower lines of (a), respectively), and C$_3$ is the ancillary qubit mediating the causal influence between C$_2$ and C$_4$. The two boxed unitaries $\mathcal{W}$ and $\mathcal{V}$ together implement the controlled operation needed to read out the correlator $\langle \sigma_i^A\sigma_j^C,\sigma_k^D\sigma_l^{B}\rangle$, where $\langle E,F\rangle$ denotes the expectation value of the product of the outcome of $E$ at time 1 and the outcome of $F$ at time 2. $\mathcal{V}$ encodes the Pauli operators $\sigma_i^A\sigma_j^C$ at time 1, while $\mathcal{W}$ encodes the Pauli operators $\sigma_k^D\sigma_l^B$ at time 2 propagated back through $\mathcal{M}$ and $\mathcal{N}$. The gates are implemented via a combination of single-spin RF pulses and internal spin-spin coupling. We choose $\mathcal{N}$ as the SWAP unitary $S$ and $\mathcal{M}$ as the partial swap $e^{-i\theta S}$, with $\theta$ controlling the swap strength.}
\label{Fig2} 
\end{figure}

\subsection*{Exploiting space-time states and time asymmetry for causal inference}
Our protocol builds on the Pseudo-Density Matrix (PDM) formalism, which generalizes the conventional density matrix by assigning a Hilbert space also to each instant in time~\cite{fitzsimons2015quantum,liu2023quantum,marletto2019theoretical,marletto2021temporal,liu2024unification,zhang2020different}. Recall that the standard density matrix for $n$ qubits can be written as a linear combination of $n$-qubit basis matrices ${\sigma}_{i}\in \{\mathbbm{1},\sigma_x,\sigma_y,\sigma_z\}^{\otimes n}$ and that the coefficients then correspond, up to a factor, to the expectation value of the given ${\sigma}_{i}$. For example, for 2 qubits $\rho= \frac{1}{4}\sum^{3}_{i_1, i_2=0} \langle \sigma_{i_1} \otimes   {\sigma}_{i_2}\rangle \, {\sigma}_{i_1} \otimes   {\sigma}_{i_2}$.  The PDM formalism uses the same expansion in terms of Paulis to represent experiments at several times. For example, for $n=1$, one may have measured $\sigma_{x}$ at time 1 and $\sigma_{z}$ at time 2 on the same qubit, with associated (correlation) expectation value $\langle \sigma_{x_1},  {\sigma}_{z_2} \rangle:=p(1,1)+p(-1,-1)-p(-1,1)-p(1,-1)$, where `,' is used instead of `$\otimes$' to indicate we may not be dealing with the two systems at the same time. The PDM $R$ for either one qubit at 2 times (cause-effect structure) or two qubits at one time then has the form $R= \frac{1}{4}\sum^{3}_{i_1, i_2=0} \langle \sigma_{i_1} ,   {\sigma}_{i_2}\rangle \, {\sigma}_{i_1} \otimes   {\sigma}_{i_2}$. More generally, $n$ qubits at two times has the same PDM form as $2n$ qubits at a single time:
\begin{equation}\label{eq: defPDM}
    R=\frac{1}{2^{2n}}\sum^{4^n-1}_{i_1, i_2=0} \langle \sigma_{i_1},  {\sigma}_{i_2} \rangle \, {\sigma}_{i_1} \otimes   {\sigma}_{i_2},
\end{equation}
where ${\sigma}_{i_\alpha}\in \{\mathbbm{1},\sigma_x,\sigma_y,\sigma_z\}^{\otimes n}$ is an $n$-qubit Pauli matrix with $\alpha=1,2$ and $\langle \sigma_{i_1},  \sigma_{i_2} \rangle$ denotes the expectation value of the product of observables $\sigma_{i_1}$ and $\sigma_{i_2}$ each on sets of $n$ qubits. We shall here mainly discuss $n=2$ and either one or two times. Thus, we have 4 different 2-dimensional Hilbert spaces $A, B, C, D$ (depicted in Fig. \ref{Fig2}(a)) and a PDM $R_{ABCD}$ on the total $2^4$-dimensional Hilbert space. 

The expectation values of Eq.\eqref{eq: defPDM} can importantly be determined via coarse grained measurements. Coarse-grained measurements here mean measurements whose elements are sums of (orthogonal) fine-grained projectors. As an example $\langle \sigma_{x} \otimes{\sigma}_{x}\rangle=\langle (\ket{+}\bra{+}\otimes \ket{+}\bra{+}+\ket{-}\bra{-}\otimes \ket{-}\bra{-})-(\ket{+}\bra{+}\otimes \ket{-}\bra{-}+\ket{-}\bra{-}\otimes \ket{+}\bra{+})  \rangle $ where the two coarse-grained projectors that are the measurement elements are the sums in brackets, $(\cdot)$. Such coarse-grained measurements are less disturbing than fine-grained measurements as they leave non-trivial subspaces of states invariant. Such measurements can be implemented by unitarily interacting the 2-qubit system of interest with a single probe qubit followed by a standard fine-grained projective measurement on that probe, as we shall discuss more later.

Defining the PDM measurement updates in terms of coarse-grained measurements leads to a neat closed form of the PDM. Suppose the dynamics is a combination of the coarse-grained measurements with the standard L\"uder projection update rule that $\rho\rightarrow \Pi \rho \Pi/\Tr(\rho \Pi)$ and a channel $\mathcal{N}$ between times 1 and 2, then the PDM has a closed-form expression as~\cite{liu2023quantum}
\begin{align}
\label{eq:PDMclosedform}
    R_{12}= \frac{1}{2}\{\rho_1 \otimes \id_2, M\},
\end{align}
where $M= \sum_{i,j} \ket{i}\bra{j} \otimes \mathcal{N}(\ket{j}\bra{i})$ is the Jamio\l{}kowski matrix associated with the channel $\mathcal{N}$ and \{. , .\} is the anticommutator. (The Choi matrix is $M^{T_B}$ where $T_B$ means transpose on $B$). We shall generally use numbered subscripts to denote the Hilbert spaces as in Eq.~\eqref{eq:PDMclosedform} when it is known that there are two times involved, and use letters as subscripts more generally, e.g.\ $R_{AB}$.

Reduced states are defined via tracing out, like for the normal density matrix, such that e.g.\ $R_{AB}=\Tr_{CD}R_{ABCD}$ is the reduced state on $A$ and $B$. Experimentally, determining $R_{AB}$ only requires the measurement data on $A$ and $B$ as with normal density matrices, in line with the PDM definition of Eq.~\eqref{eq: defPDM}.

The presence of negativity in $R_{AB}$ certifies that the measurements labeled by $A$ and $B$ correspond to two distinct times (cause-effect mechanism) rather than a single time (common-cause mechanism). If a PDM has negative eigenvalues there must be multiple times involved since the standard, single-time, density matrix is positive (assuming sufficient experimental data was taken to eliminate apparent negativity due to experimental error)~\cite{fitzsimons2015quantum}. In practice, the high signal-to-noise ratio of the NMR ensemble and the high gate fidelities keep experimental errors small. We treat a measured $f(R_{AB})$ as consistent with zero (and hence as showing no negativity) when it lies within one standard deviation of zero, and as indicating genuine negativity only when it exceeds zero by at least one standard deviation; this criterion guards against spurious negativity from experimental noise. As a check, at the parameter values where zero negativity is theoretically expected (e.g., $\lambda=0$ or $\theta=0,\pi$), the measured $f(R_{AB})$ is indeed consistent with zero. Wishing to determine whether the $A$ and $B$ data is associated with the same time or different times (cause-effect structure), we thus evaluate the negativity (the presence of negative eigenvalues) of $R_{AB}$ as the first step of the data analysis. The negativity of $R$ (or some other square matrix) can be defined as~\cite{fitzsimons2015quantum}
\begin{equation}\label{Eq5}
f(R):=\Tr \sqrt{RR^\dag} - \Tr \, R .
\end{equation}
$f(R)>0$ when $R$ has negative eigenvalues, and $f(R)=0$ when $R$ is positive semidefinite~\cite{fitzsimons2015quantum}. The first step is thus to check whether $f(R_{AB})>0$. If $f(R_{AB})=0$ the data is compatible with two measurements at a single time which is case 3 in FIG.\ref{fig:Fig1}, the common cause case. If $f(R_{AB})>0$ there are multiple, in our case 2, times involved, $t_1$ and $t_2$, and we want to determine whether $R_{AB}$ from the experimental data is $R_{12}$ or its time-reversal~\cite{liu2023inferring,song2023causal,PRXQuantum.4.020334} 
\begin{equation}\label{eq:reversePDM}
R_{21}:= S \, R_{12} \, S^\dag, 
\end{equation}
where $S=\sum_{i,j}\ket{j,i}\bra{i,j}$ denotes the operator that swaps two Hilbert spaces. 

To determine the temporal ordering, we focus on whether the dynamics of the PDM, and/or of the time-reversed PDM are allowed quantum dynamics. The dynamics implied by a given PDM can be represented as the Choi matrix representation of the channel: 
$M^{T_B}$ (with $M$ defined around Eq.\eqref{eq:PDMclosedform})~\cite{liu2023quantum,liu2023inferring,supp}. For example, in the special case where the initial state $\rho_A=\Tr_B R_{AB}$ is full rank, $M_{AB}= 2\int^{\infty}_{0} e^{-t \rho_A \otimes \id_B }\, R_{AB}\, e^{-t\rho_A\otimes \id_B} dt$~\cite{liu2023quantum,liu2023inferring,supp}. (For the case of not full rank see the code in Sec.27, SM~\cite{supp}.) It is known that $M^{T_B}$ is positive ($f(M^{T_B})=0$) exactly if the channel is allowed (completely positive). Thus, if $f(M^{T_B}_{AB})>0$ but $f(M^{T_B}_{BA})=0$ then we have compatibility with $R_{AB}$ being $R_{21}$ (i.e.\ the structure $B\rightarrow A$) but not with $R_{AB}$ being $R_{12}$ ($A\rightarrow B$) and vice versa.

The following table summarises the assignment of causal structure compatibility as a function of the negativity $f(\cdot)$ of the experimentally determined matrices $R_{AB}$ and $M^{T_B}_{AB}$~\cite{liu2023quantum}. 
\begin{tabular}{|c|c|c|c|}
  \hline
  $f(R_{AB})$ & $f(M^{T_B}_{AB})$ & $f(M^{T_B}_{BA})$ & $R_{AB}$ is compatible with  \\
  \hline
 0 & Any & Any & Common Cause \\
  \hline
  $>0$ & 0 &   $>0$ & $ A \rightarrow B $ \\
  \hline
  $>0$ &   $>0$ & 0  &   $B \rightarrow A $ \\
  \hline
    $>0$ & 0 & 0 &  $ A \rightarrow B $ or  $ B \rightarrow A $\\
  \hline
      $>0$ & $>0$ & $>0$ & Mixture\\
  \hline
\end{tabular}

The causal inference protocol is thus to determine $R_{AB}$ from measurements,  extract the $A-B$ dynamics $M^{T_B}_{AB}$, and the time reversed
dynamics $M^{T_B}_{BA}$, evaluate the negativities $f(.)$ and then use the above table to assign compatibility with a causal structure(s).

We demonstrate the protocol in four experiments. The first two experiments concern two qualitatively different realizations of the causal structure $A\rightarrow B$, with a unitary and fully decohering channel, respectively. The third experiment concerns the common-cause structure, and the fourth the mixture of the common-cause and cause-effect structures. The first two are mainly described in the main body, and further in SM Sections S2, S3 and SM Figs. S2-S7. The last two are described in the SM sections S4, S5 and SM Figs. S8-S12~\cite{supp}. The experiments were implemented on a four-qubit NMR quantum processor~\cite{xi2024experimental,Huangkeyi2024,Longxy2022,lin2024hardware}, depicted in Fig.~\ref{Fig2}(b). In such a processor the dynamical evolution can be controlled via a combination of single-spin radio-frequency (RF) pulses and internal spin-spin coupling (See S2 of SM~\cite{supp}.

\subsection*{Experiment for $A\rightarrow B$, unitary case}
We firstly undertook an experiment to detect compatibility with the {\em a priori} unknown causal structure $A\rightarrow B$ whereby the preparation of $A$ impacts the measurements of $B$ and not vice-versa (as depicted in Fig.~\ref{Fig2}(a)). 

We implement the causal structure $A\rightarrow B$ via the following unitary circuit. We set the initial state of the four qubit quantum processor as $\ket{+}\bra{+}\otimes\rho_A\otimes\ket{00}\bra{00}$, where $\rho_A=(1-\lambda)\frac{\mathbb{I}}{2}+\lambda\ket{+}\langle+|$, with $\lambda$ being the polarization parameter (a full-rank state). To avoid bi-directional influence an intermediate spin C$_3$ is used (see Fig.~\ref{Fig2} (a))~\cite{beckman2001causal}. For experimental convenience, we prepare the ancillary system (spin C$_3$) and the lower-most system (spin C$_4$) in the $\ket{00}$ state.
The parameter $\lambda$ is experimentally controlled via the rotation parameter $\alpha=\arccos(\lambda)$ in Fig.~\ref{Fig2}(c), which is combined with dephasing induced by the gradient-field pulse Gz to prepare $\rho_A$.
$\mathcal{N}$ of Fig.~\ref{Fig2}(a) is chosen as the 2-qubit swap unitary $S=\ket{00}\bra{00}+\ket{10}\bra{01}+\ket{01}\bra{10}+\ket{11}\bra{11}$ and $\mathcal{M}$ as the partial swap $e^{-i\theta S}$. $\theta$ determines the strength of the swap. When $\theta=0, \pi$, the gate acts as the identity (i.e., no swap occurs). If $\theta=\pi/2$, the gate implements a complete swap (up to a global phase).

\begin{figure*}
    \includegraphics[width=1\linewidth]{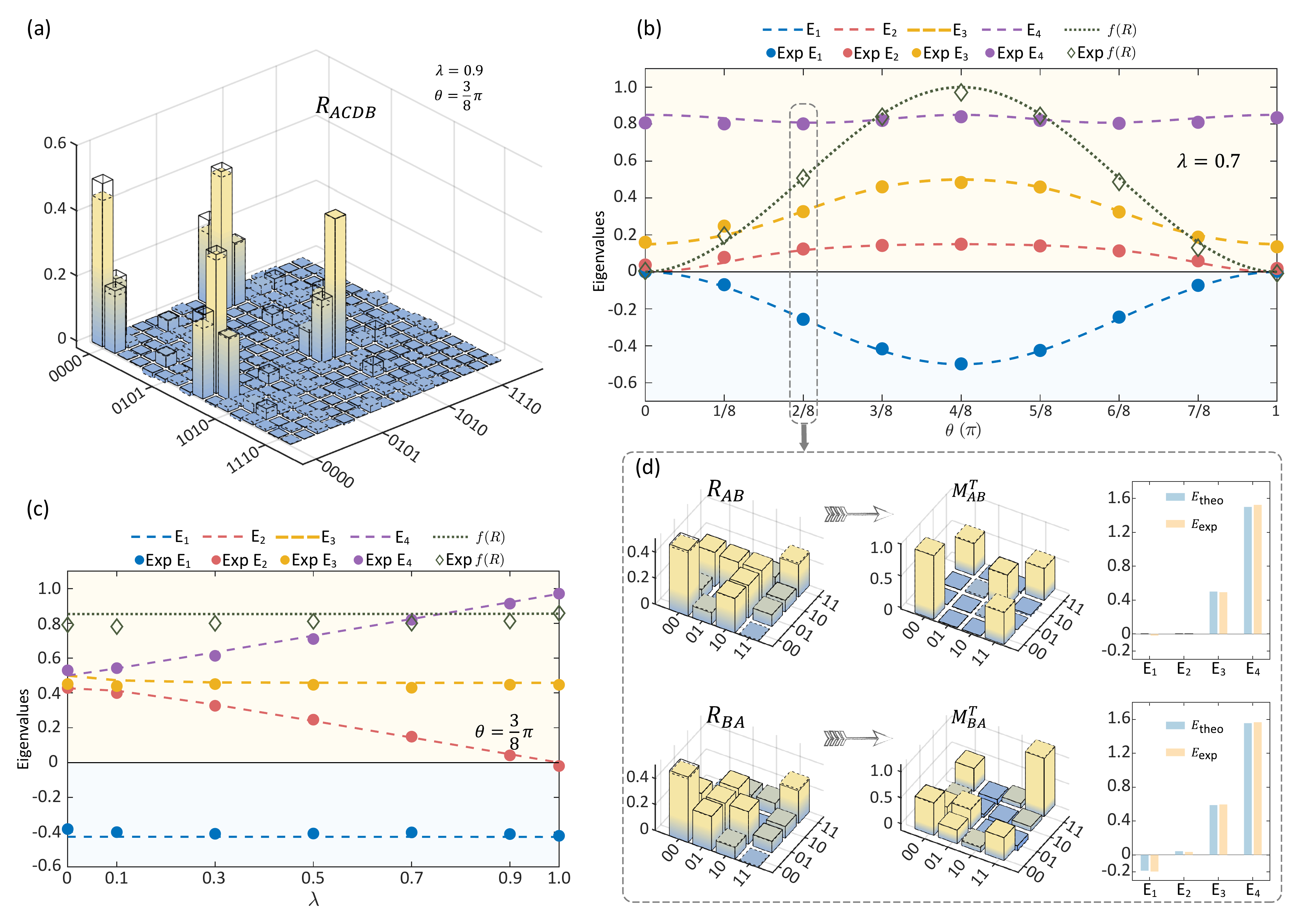}
    \caption{{\bf Experimentally acquired data for causal inference highly consistent with theoretical predictions.} (a) The full PDM $R_{ACDB}$ for $\lambda=0.9$ and $\theta=3\pi/8$.  The solid bars represent the theoretical prediction, while the color bars with dashed lines indicate the experimentally acquired values. (b) The 4 eigenvalues of the reduced PDM $R_{AB}$ for different $\theta$. $E_1$ to $E_4$ are the theoretical predictions whilst Exp $E_1$ to Exp $E_4$ are inferred from experimental data. The negativity of the PDM $f(R)$ is also shown similarly.
    (c) is the same as (b) except for the x-axis being the initial state parameter $\lambda$. In (b)  $\lambda=0.7$, and in (c) the $\theta=3\pi/8$; the error bars are within the size of the marker symbols. 
   (d) The experimentally measured reduced PDMs $R_{AB}$ and $R_{BA}$ are shown together with the corresponding inferred Choi matrices $M^{T_B}$ (denoted as $M^T$ for graphical simplicity). The eigenvalues of the Choi matrices are shown next to their experimental predictions for $\lambda=0.7$ and $\theta=2\pi/8$. The absence or presence of negative eigenvalues in such Choi matrices is used in determining compatibility with given causal structures.} 
\label{Fig3} 
\end{figure*}

To determine $R_{AB}$ via coarse-grained projective measurements, we use what is known as a quantum scattering circuit~\cite{miquel2002interpretation,souza2011scattering,Niexf2020otoc} to obtain the expectation values of Eq.~\eqref{eq: defPDM} that determine $R_{AB}$. 
One qubit acts a probe, interacting with the 2 qubits of interest and then undergoing a projective measurement, as depicted in Fig.~\ref{Fig2}(c). We prove in SM Sec.S1~\cite{supp} that these circuits indeed give the expectation values of Eq.~\eqref{eq: defPDM}:
\begin{align}\label{eq:sigmaz}
    \braket{\sigma_z}_{\text{probe}} 
    =& \braket{\sigma_i\otimes\sigma_j, \sigma_k \otimes\sigma_l } \nonumber\\
=&\braket{\Pi^{(i,j)}_+, \Pi^{(k,l)}_+}+\braket{\Pi^{(i,j)}_-, \Pi^{(k,l)}_-} \nonumber\\
    &-(\braket{\Pi^{(i,j)}_-, \Pi^{(k,l)}_+}+\braket{\Pi^{(i,j)}_+, \Pi^{(k,l)}_-}).
\end{align}
Here, $\Pi^{(i,j)}_{+}= \Pi^i_+ \otimes \Pi^j_+ +\Pi^i_- \otimes \Pi^j_- $ denotes the coarse-grained projector onto the $+1$ eigenspace of $\sigma_i\otimes\sigma_j$, $\Pi^{(i,j)}_{-}= \Pi^i_- \otimes \Pi^j_+ +\Pi^i_- \otimes \Pi^j_+ $ denotes the coarse-grained projector onto the $-1$ eigenspace of $\sigma_i\otimes\sigma_j$, with a similar definition for $\Pi^{(k,l)}_{\pm}$. The first of the four qubits (C$_1$) acts as the probe, the second (C$_2$) and fourth (C$_4$) qubits as the quantum systems of interest, and the third qubit (C$_3$) as an ancillary system. 

In order to firstly verify the experimental setup, we acquired sufficient experimental data to reconstruct the full PDM $R_{ACDB}$. (The causal inference does not require the full PDM but rather $R_{AB}$.) We measured the $\sigma_z$ expectation value of the probe qubit, $\langle \sigma_z \rangle_{ \text{probe}}$, for the complete set of Pauli operators on $A, B, C$ and $D$. As representative non-extremal values of the free parameters we chose arbitrarily $\lambda=0.9$ and $\theta=3 \pi/8$. This data enabled us to construct the experimental PDM $R_{ACDB}$, as shown in Fig.~\ref{Fig3}(a). The agreement between the experimental results and theoretical predictions is high.

We acquired experimental data to evaluate $f(R_{AB})$ required for the causal inference protocol of the above Table.  We acquired data for constructing the reduced PDM $R_{AB}$. As $R_{AB}$ was sufficient for our quantum causal inference protocol, we set $\sigma^C=\sigma^D=\id$. We collected the data on the probe system, $\langle \sigma_z \rangle_{ \text{probe}}$ for $R_{AB}$ for a range of values. Firstly, we set $\lambda=0.7$ and then systematically varied the channel parameter $\theta$ from $0$ to $\pi$. We constructed the PDM $R_{AB}$ from that data. The eigenvalues $E_{i}$ and negativity $f(R_{AB})$ of $R_{AB}$ are depicted in Fig.~\ref{Fig3}(b). We observe that $f(R_{AB})>0$ except for two extreme values $\theta=0$ and $\theta=\pi$, where the channel $\mathcal{M}$ becomes trivial, i.e, $e^{-i\theta S}=\pm\mathcal{I}$. Secondly, we varied the polarization parameter $\lambda$ while fixing $\theta$ at $3\pi/8$. The eigenvalues $E_{i}$ and negativity $f(R)$ of the experimentally constructed $R_{AB}$ are depicted in Fig.~\ref{Fig3}(c). Again $f(R_{AB})>0$. Thus, from this step, we conclude that the data concerns two times and that the ordering of those two times needs to be determined.

The next step of the protocol is to determine $f(M^{T_B})$. From the above $R_{AB}$  (which was for $\lambda=0.7$ and $\theta=\pi/4$) we calculate $R_{BA}=SR_{AB}S^{\dagger}$  and the respective Choi matrices $M^{T_B}_{AB}$ and $M^{T_B}_{BA}$. We evaluated the positivity of the Choi matrices, finding  $f(M_{AB}^{T_B})=0$ (positive) and $f(M_{BA}^{T_B})>0$ (negative). Thus the ordering is determined as compatible with $A\rightarrow B$, successfully concluding the causal inference protocol. 

The experimental results and outcomes of the analysis are displayed in Fig.~\ref{Fig3}(d), showing high agreement between theoretical predictions and experiment. Further details are given in the SM Sec. S2~\cite{supp}.

\begin{figure}
    \includegraphics[width=1\linewidth]{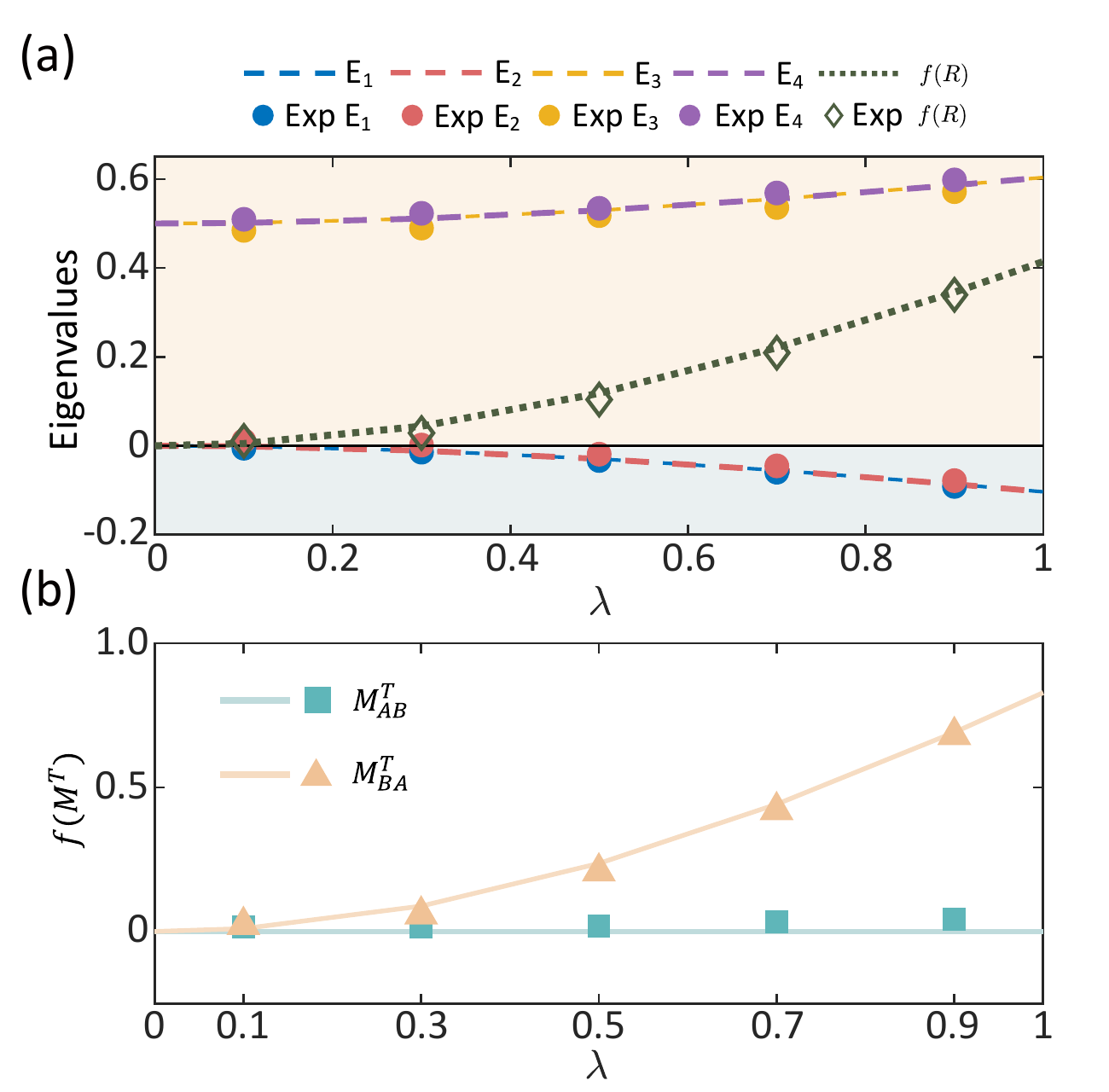}
    \caption{{\bf Causal inference for fully decohering channel.} (a) The experimental results of the eigenvalues of $R_{AB}$ and the values of $f(R)$ vary with $\lambda$. The circles and diamonds indicate eigenvalues and $f(R)$ data points obtained from the experiment, respectively, while the dotted lines indicate the theoretical predictions. As $\rho_a$ transitions from the maximally mixed state to the quantum state $\ket{+}\bra{+}$, the magnitude of the negative eigenvalues correspondingly increases. When $\rho_a$ is in the $\ket{+}\bra{+}$ state and passes through the decoherence channel, the causal strength from A to B is maximized. Conversely, when $\rho_a$ reaches the maximally mixed state, the causal correlation between A and B becomes undetectable. The error bars are within the size of the marker symbols. (b) Experimental results of $f(M^{T_B}_{BA})$ and $f(M^{T_B}_{AB})$ are shown as squares and triangles for different $\lambda$. Theoretical predictions are shown as lines.}
\label{Fig4} 
\end{figure}

\subsection*{Experiment for $A \rightarrow B$, fully decohering channel}
We further applied our causal inference protocol to the cause-effect mechanism where the functional relation of the quantum systems includes a fully decohering channel $\mathcal{L}(\cdot)$. This probed whether employing coarse-grained projective measurements alone for causal inference also works when the evolution is, in a sense, classical. Specifically $\mathcal{N}$ of Fig.~\ref{Fig2}(a) was taken as the fully decohering measure-and-prepare channel, i.e., measuring $C_1$ and then preparing the results on $C_4$:
\begin{align}
    \mathcal{L}(\rho_{C_1C_4})=\langle 0|\rho_{C_1}\ket{0} \ket{00}_{C_1C_4}\bra{00}
+\langle1|\rho_{C_1}\ket{1}\ket{11}_{C_1C_4}\langle11|.
\end{align}
$\mathcal{M}$ of Fig.~\ref{Fig2}(a) is a swap $S$ swapping $C_2$ and $C_4$. (The corresponding circuit for constructing the PDM $R_{AB}$ and the experimental details are given in the SM Sec. S3~\cite{supp}.) In this case, a standard circuit was used rather than a scattering circuit. Since the channel $\mathcal{N}$ is itself a measure-and-prepare operation that already destroys coherence, the advantage of the scattering circuit is less critical. The standard circuit avoids the need for an additional probe qubit, fitting within the four-qubit constraint of our NMR processor. We note that the scattering circuit approach could equally be applied here if more qubits were available, and our protocol is fully scalable in this regard. The initial state of the system is $\rho_A\otimes\vert0\rangle\langle0\vert\otimes\frac{\mathbb{I}}{2}\otimes\vert0\rangle\langle0\vert$, where again $\rho_A=(1-\lambda)\frac{\mathbb{I}}{2}+\lambda\ket{+}\langle+|$.

We gathered data to determine $f(R_{AB})$. Upon obtaining the data $\langle \sigma^A_i\sigma^B_j\rangle$ for various $\lambda$, we experimentally constructed the PDMs $R_{AB}$ and analysed it. The results are shown in Fig.~\ref{Fig4}.  We observed negative eigenvalues in $R_{AB}$ except for $\lambda=0$, as depicted in Fig.~\ref{Fig4}(a). The negativity ruled out the pure common-cause structure. We then proceeded similarly to above to determine the time ordering via $f(M^{T_B})$. We extracted the Choi matrices $M_{AB}^T$ and $M_{BA}^T$ from $R_{AB}$ and $R_{BA}$, respectively and calculated their negativities. The results are shown in Fig.~\ref{Fig4}(b). 
Also here $f(M_{AB}^T)=0$ (positive) and $f(M_{BA}^T)>0$ (negative), such that $A$ must have preceded $B$. Combining the above, we concluded that the causal structure is compatible with $A\to B$, as desired, for initial state parameter $\lambda \neq 0$.

The protocol fails as expected for $\lambda=0$, the case corresponding to the classical limit of no initial coherence. For $\lambda=0$, our protocol incorrectly determines compatibility with the common cause structure since $f(R)=0$. This is because the setup fully returns to the classical case, with the initial state having no coherence in the computational basis. This is consistent with the hypothesis that in classical probability theory, we {\em cannot} in general distinguish the causal structure purely based on observational data. The difference between the $\lambda=0$ case and the other cases of initial coherence can be viewed as confirming the usefulness of state coherence in inferring causal structures, at least for the case of Pauli measurements.

\section*{Discussion}
We have conducted high-precision NMR experiments that, together with analysis of the negativity properties of pseudo-density matrices, demonstrate the experimental discrimination between several possible causal structures.
The experiments solely employ coarse-grained measurements implemented via ancilla systems using so-called scattering circuits.  

The successful results demonstrate experimentally an extension of our ability to infer causal structure without invasive interventions. The experiments and analysis go beyond previous experiments in this direction by distinguishing a wider range of scenarios, in particular scenarios with combinations of temporal and spatial correlations in bipartite systems at two times.   

A key limitation of the current results concerns distinguishing structures 4 and 5 in Fig.~\ref{fig:Fig1}.
This motivates seeking a deeper quantitative understanding of the arrow of time when there are correlations in the initial state. Another limitation appeared in the case of a diagonal initial state ($\lambda=0$) with a fully decohering channel. An interesting question is whether other measurement choices may reveal the causal structure in that case.

The approach can be developed in related directions. The experimental data specifies the functional relation $M_{AB}$, describing the intermediate dynamics between the measurements. This approach may thus lead to new types of multiqubit channel tomography protocols based on coarse-grained measurements alone.
The potential advantage is not in the scaling, but in that preparing several input states is then not necessary for channel tomography. Instead several initial Pauli measurements are required. However, the scattering circuit approach to coarse-grained measurements makes the initial Pauli measurement easier, replacing the initial Pauli measurement with an interaction with the probe qubit. Further, the approach here allows for correlations between the initial state and the environment, suggesting the approach can be used to design witnesses of non-Markovianity.

\section*{METHODS}
\subsection*{Experimental implementation of causal inference protocol}
We experimentally demonstrate our quantum causal inference protocol, which relies only on minimally invasive coarse-grained projective measurements without resorting to interventions that reset the system, on the NMR platform by inferring a cause-effect mechanism.
As shown in the left panel of Fig.~\ref{Fig2} (a), the directed acyclic graph represents the cause-effect mechanism in which only $A$ can influence $B$ and there is no shared cause (initial correlation). 

This causal structure can be modeled by the circuit depicted in the right panel of Fig.~\ref{Fig2} (a). In the experiment, four qubits are used. The physical qubit C$_2$ carries events $A$ (time 1) and $D$ (time 2), while C$_4$ carries events $C$ (time 1) and $B$ (time 2). The ancillary qubit C$_3$ mediates the causal influence: C$_2$ first interacts with C$_3$ via the channel $\mathcal{N}$ (a SWAP gate), and then C$_3$ interacts with C$_4$ via the channel $\mathcal{M}$ (a partial SWAP). This intermediary architecture ensures that C$_4$ cannot back-act on C$_2$, physically enforcing the exclusion of causal loops assumed in our protocol. In addition, C$_1$ serves as the probe qubit in the scattering circuit, enabling non-invasive extraction of expectation values without directly measuring the system qubits. Further, we label quantum systems at different times as $A, B, C$, and $D$, and there are no correlations between $A$ and $C$.  
To facilitate the experiment, we consider those quantum systems as qubits and set the channels 
\begin{align}\label{eq:channels}
   \mathcal{N}(\cdot)=S(\cdot)S^{\dag} \ \text{and} \ \mathcal{M}(\cdot)=e^{-i\theta S}(\cdot)e^{i\theta S} 
\end{align}
 where $S$ denotes the SWAP operator and $\theta$ is a parameter, and the initial state as follows:
 \begin{align}\label{eq:initialAC}
      \rho_{AC}=[ \lambda\ket{+}\bra{+} +(1-\lambda)\mathbb{I}/2 ]\otimes\ket{0}\bra{0}.
 \end{align}
Here, $\lambda$ is the polarization of $A$.

Our quantum causal inference experiment relies on constructing the PDM $R_{AB}$, which depends on the coarse-grained measurement of the expectation values $\langle \sigma_i^A, \sigma_j^B \rangle$. 

To minimize the disturbance on the system caused by the measurement process, we introduce the quantum scattering circuit~\cite{miquel2002interpretation,PhysRevA.105.L030402}, as illustrated in Fig.~\ref{Fig2} (c), to obtain the expectation value $\langle \sigma_i^A, \sigma_j^B \rangle$. The circuit is equipped with a probe qubit that interacts with the quantum system. In the SM, we show that
 \begin{align}
 \langle \sigma_z \rangle_{ \text{probe}}=  \langle \sigma_i^A, \sigma_j^B \rangle :=   \langle \sigma_i^A \id^C, \id^D \sigma_j^B \rangle .
 \end{align}
 In this way, one can obtain $\langle \sigma_i^A, \sigma_j^B \rangle$  by measuring the probe qubit thus circumventing the need for repeated direct measurements of the quantum systems of interest. 
In addition to the three qubits constructing the cause-effect mechanism, four qubits are required experimentally.

The NMR platform is employed to implement the above process. Our experiments are conducted on a Bruker 600 MHz spectrometer at room temperature.  The NMR four-qubit quantum processor comprises four C nuclear spins within $\text{C}$-labeled trans-crotonic acid dissolved in $d_6$ acetone~\cite{Longxy2022,xi2024experimental} (See also Fig. S2 in SM~\cite{supp}), see Fig.~\ref{Fig2}(b). The first spin C$_1$ acts as the probe qubit, spins C$_{2,4}$ comprise the two quantum systems of interest, and C$_3$ constitutes the ancillary qubit, which intermediates the causal influence between the spins  C$_{2,4}$.
The quantum circuit in Fig.~\ref{Fig2}(c) can be divided into three segments: 
(1) initialization; (2) unitary evolution; (3) results readout. 
Next, we introduce each segment of the quantum circuit in detail.

(1) Initialization. The four-qubit system started from the pseudo-pure state $\ket{0000}\bra{0000}$. For experimental convenience, the ancillary qubit and $C$ remain unchanged in this stage.   The mixed state $\rho_A=\lambda\ket{+}\bra{+} +(1-\lambda)\mathbb{I}/2 $ is prepared using two single-qubit rotations along the y-axis, inserted with a gradient-field pulse~\cite{lin2024hardware}. 
The rotation angles correspond to $\alpha=\text{arccos}\lambda$ and $\pi/2$, respectively. Further, a Hadamard gate is applied on the probe qubit to transfer it to $\vert+\rangle$. Therefore, the four-qubit system is initialized to be $\rho_0=\ket{+}\bra{+}\otimes\rho_{A}\otimes\ket{00}\bra{00}$.

(2) Unitary evolution.  The experimental circuit is designed to realize the unitary $\mathcal{U}_\text{c}$ $=\vert0\rangle\langle0\vert\otimes\mathbbm{I}+\vert1\rangle\langle1\vert\otimes \mathcal{V}\mathcal{W}$ on the four-qubit system.
The explicit forms of unitaries $\mathcal{V}$ and $\mathcal{W}$ are determined by the quantum channels, $\mathcal{N}$ and $\mathcal{M}$, as well as the time correlator $\langle \sigma_i^A, \sigma_l^B \rangle$ we aim to measure. 
Let $U_\mathcal{N}$ and $U_\mathcal{M}$ be the corresponding unitaries when $\mathcal{N}$ and $\mathcal{M}$ are unitary channels. In Fig.~\ref{Fig2}(c), the quantum circuit is designed to obtain $\langle \sigma_i^A\sigma_j^C,\sigma_k^D\sigma_l^{B}\rangle$.    For this specific choice of the Pauli operators, $\mathcal{W}=(U_\mathcal{M}U_\mathcal{N})^\dagger\sigma_k^2\sigma_l^4 (U_\mathcal{M}U_\mathcal{N})$ and $\mathcal{V}=\sigma_i^2\sigma_j^4$.
To obtain $\langle \sigma_i^A ,\sigma_l^B \rangle$, one set $\sigma^C=\sigma^D=\id$.
All the single-qubit rotations and two-qubit gates are fundamental quantum operations that can be easily implemented in most physical systems. For the NMR experimental apparatus, experimental control accuracy is further improved by creating a shaped pulse through gradient-based optimization~\cite{KHANEJA2005296}.

(3) Results readout. After the evolution mentioned above, the expectation value $\langle \sigma_i^A\sigma_j^C,\sigma_k^D\sigma_l^{B}\rangle$ is encoded in the coherence term of the probe qubit.  Subsequent application of a Hadamard gate allows direct extraction of the PDM element by measuring  $\langle \sigma_z^1\rangle$ of the probe qubit. In this way, we can construct the PDM $R_{ACDB}$, and the reduced PDM $R_{AB}$ when the expectation values  $\langle \sigma_i^A,\sigma_l^{B}\rangle$ are selected.
To construct $R_{AB}$ and $R_{ACDB}$, the experiment must be repeated 16 and 256 times, respectively, each time with a refreshed $\mathcal{U}_\text{c}$ corresponding to the specific elements.

\textbf{Acknowledgements:}
This work is supported by the National Key Research and Development Program of China (2019YFA0308100), and the National Natural Science Foundation of China (Grants No. 12050410246, 1200509, 12050410245, 12574543, 12341101), Science, Technology and Innovation Commission of Shenzhen Municipal
City (JCYJ20200109140803865), Guangdong Innovative and Entrepreneurial Research Team Program (2019ZT08C044), Guangdong Provincial Key Laboratory (2019B121203002), City University of Hong Kong (Project No. 9610623), The Pearl River Talent Recruitment Program (2019QN01X298), Guangdong Provincial Quantum Science Strategic Initiative (GDZX2303001, GDZX2403004 and GDZX2200001), and the National Research Foundation through the NRF Investigatorship on Quantum-Enhanced Agents (Grant No. NRF-NRFI09-0010) and the National Quantum Office, hosted in A*STAR, under its Centre for Quans Funding Initiative (S24Q2d0009), the Singapore Ministry of Education Tier 1 Grant RT4/23 and RG77/22 (S), the FQXi R-710-000-146-720 Grant “Are quantum agents more energetically efficient at making predictions?” from the Foundational Questions Institute, Fetzer Franklin Fund (a donor-advised fund of Silicon Valley Community Foundation). QC is funded within the QuantERA II Programme that has received funding from the European Union’s Horizon 2020 research and innovation programme under Grant Agreement No 101017733 (VERIqTAS). YQ is supported by the National Research Foundation, Singapore, and A*STAR under its CQT Bridging Grant and its Quantum Engineering Programme under grant NRF2021-QEP2-02-P05. V.V is supported by the Gordon and Betty Moore and Templeton Foundations.

\section*{References}
\bibliography{references_sa}

%%%%%%%%%%%%%%%%%%%%%%%%%%%%%%%%%%%%%%%%%%%%%
\section*{Declarations}

\noindent \textbf{Author contributions:}
H.L. and X.L. contributed equally to this work.
H.L. performed the experiments under the supervision of D.L.,
X.L. and O.D. carried out the theoretical derivations and numerical simulations. 
Q.C. assisted with data processing. 
X.N. contributed to the experimental design. 
All authors, H.L., X.L., Q.C.  Y.Q., V.V., X.N, O.D. and D.L. discussed the results and contributed to the development and writing of the manuscript. All authors have read and approved the manuscript. 

\noindent \textbf{Data and materials availability:}
All data needed to evaluate the conclusions in the paper are present in the paper and/or the Supplementary Materials.

\clearpage
\onecolumngrid

% Restart supplemental numbering
\setcounter{section}{0}
\setcounter{equation}{0}
\setcounter{figure}{0}
\setcounter{table}{0}

\renewcommand{\theequation}{S\arabic{equation}}
\renewcommand{\thefigure}{S\arabic{figure}}
\renewcommand{\thetable}{S\arabic{table}}

% Distinct hyperlink destinations
\renewcommand{\theHequation}{supp.\arabic{equation}}
\renewcommand{\theHfigure}{supp.\arabic{figure}}
\renewcommand{\theHtable}{supp.\arabic{table}}
\renewcommand{\theHsection}{supp.\arabic{section}}

\begin{center}
    \large\bfseries Supplemental Material
\end{center}

\section{Implementing Coarse-Grained Measurements via Quantum Scattering Circuit}
\label{sec: scatteringcircuit}

\begin{figure}[b]
    \includegraphics[width=0.7\linewidth]{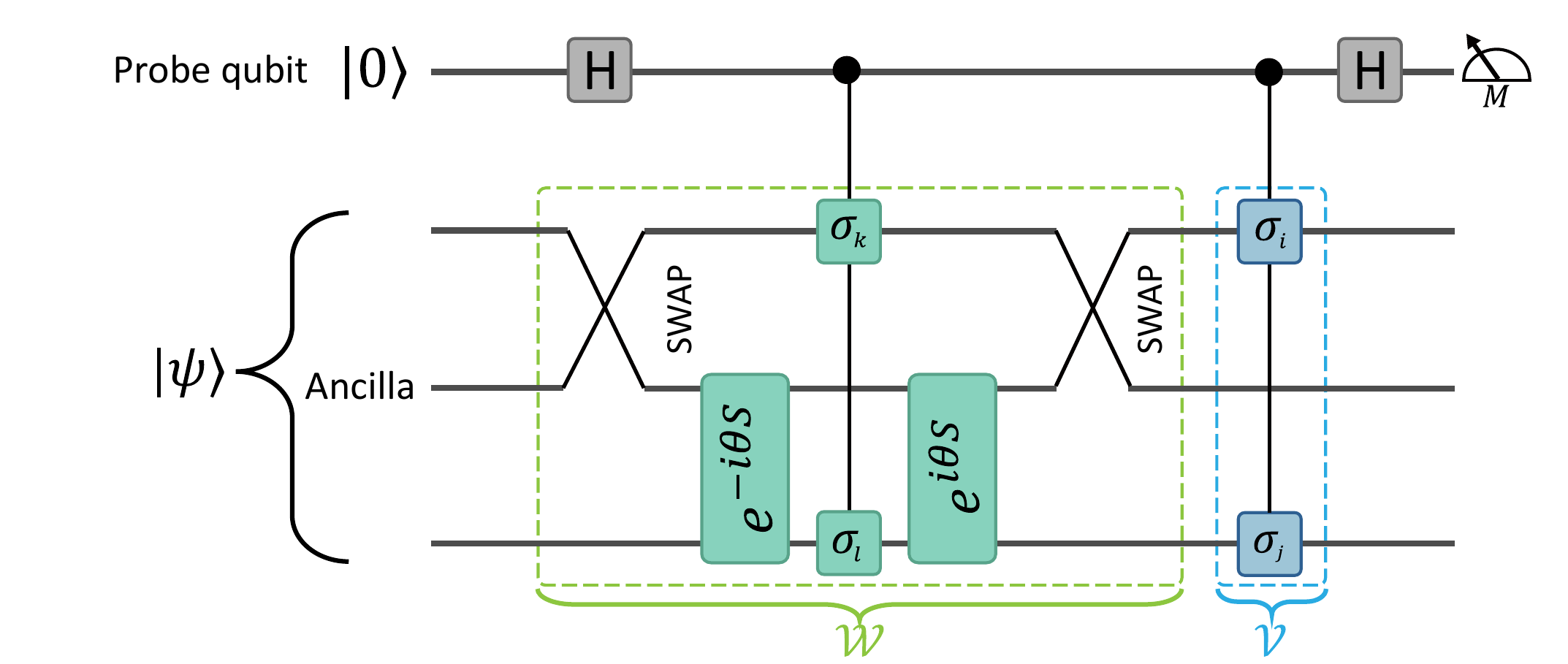}
    \caption{Experimental quantum circuit for measuring the expectation values $\langle \sigma_i\otimes\sigma_j,\sigma_k\otimes\sigma_l\rangle$.}
\label{fig: S1circuit} 
\end{figure}

The 2-time PDM employed to infer the quantum causal structure is constructed from the two-time correlators and the corresponding Pauli operators. To derive a closed-form of the 2-time PDM, the coarse-grained measurements are used to obtain the two-time correlators. This section shows that one can obtain the same values of the two-time correlators by the quantum scattering circuit shown in Fig.~\ref{fig: S1circuit}. The demonstration is divided into the following two steps.

Firstly, we show that the expectation value of $\sigma_z$ on the output of the probe qubit, $\langle \sigma_z \rangle_{ \text{probe}}$, equals the two-time correlator. In Fig.~\ref{fig: S1circuit}, the probe qubit is in the state of $\ket{0}$ and the other three qubits are in the state of $\ket{\psi}$. The overall state undergoes the following evolution:
\begin{align}
   \ket{0}\ket{\psi} &\to \ket{+} \ket{\psi} \to \frac{1}{\sqrt{2}}(\ket{0}\ket{\psi}+ \ket{1} \mathcal{W}\ket{\psi} \to  \frac{1}{\sqrt{2}}(\ket{0}\ket{\psi}+ \ket{1} \mathcal{V}\mathcal{W}\ket{\psi} ) \nonumber\\
   &\to \frac{1}{2}(\ket{0}(\mathbb{I}+\mathcal{V}\mathcal{W})\ket{\psi} +\ket{1}(\mathbb{I}-\mathcal{V}\mathcal{W})\ket{\psi} ) ,
\end{align}
where $\mathcal{W}= S_{A,Ancilla} \circ e^{i \theta S_{Ancilla,C}} \circ 
 ( \sigma_k^A \otimes \sigma^C_l ) \circ   e^{ -i \theta S_{Ancilla,C}} \circ S_{A,Ancilla}:= U^\dag ( \sigma_k^A \otimes \sigma^C_l ) U$ and $\mathcal{V}=\sigma^{A}_i \otimes \sigma^C_j$. Therefore, the value of $\langle \sigma_z \rangle_{ \text{probe}}$  at the end is given by
\begin{align}\label{eq:SI-sigmaz}
    \braket{\sigma_z}_{probe} &= \frac{1}{4} \bra{\psi}(2\mathbb{I} + \mathcal{V}\mathcal{W}+\mathcal{W}^\dag\mathcal{V}^\dag)\ket{\psi}-\frac{1}{4} \bra{\psi}(2\mathbb{I} -\mathcal{V}\mathcal{W}-\mathcal{W}^\dag\mathcal{V}^\dag)\ket{\psi} \nonumber\\
    &= \frac{1}{2} \bra{\psi}(\mathcal{V}\mathcal{W}+\mathcal{W}^\dag\mathcal{V}^\dag)\ket{\psi} \nonumber\\
    &= \tr\left[(\sigma_k^A \otimes \sigma^C_l )  U (\sigma^A_i \otimes \sigma^C_j)U^\dag \rho   \right] \nonumber\\
    &= \braket{\sigma_i\otimes\sigma_j, \sigma_k \otimes\sigma_l },
\end{align}
where here  $\rho=\ket{\psi}\bra{\psi}$.
%$\rho=\ket{0}\bra{0} \otimes \ket{\psi}\bra{\psi}$. 
The generalization of the above calculation to mixed states is straightforward.

Secondly, we show that the quantity $\braket{\sigma_z}_{probe}$ in Eq.~\eqref{eq:SI-sigmaz} equals the two-time correlators obtained by coarse-grained measurements. In particular,
\begin{align}
    \braket{\sigma_i\otimes\sigma_j, \sigma_k \otimes\sigma_l } &= \braket{\Pi^{(i,j)}_+-\Pi^{(i,j)}_-, \Pi^{(k,l)}_+-\Pi^{(k,l)}_-} \nonumber\\
    &=\braket{\Pi^{(i,j)}_+, \Pi^{(k,l)}_+}+\braket{\Pi^{(i,j)}_-, \Pi^{(k,l)}_-}-(\braket{\Pi^{(i,j)}_-, \Pi^{(k,l)}_+}+\braket{\Pi^{(i,j)}_+, \Pi^{(k,l)}_-}),
\end{align}
where $\Pi^{(i,j)}_{+}= \Pi^i_+ \otimes \Pi^j_+ +\Pi^i_- \otimes \Pi^j_- $, $\Pi^{(i,j)}_{-}= \Pi^i_- \otimes \Pi^j_+ +\Pi^i_- \otimes \Pi^j_+ $ and similarly for $\Pi^{(k,l)}_{\pm}$.

In conclusion, we have demonstrated that the quantum scattering circuit can be employed to implement the coarse-grained measurement scheme.

\section{Experimental Details for `Experiment for $A \to B$, Unitary Case'}\label{metho}

\subsection{Characterization, Pseudo-pure State Preparation, and Initialization}

We provide the experimental details for the section `Experiment for $A\to B$, unitary case', including Characterization, Pseudo-pure state preparation, and initialization. 

\begin{figure}[b]
\centering
\includegraphics[width=0.7\linewidth]{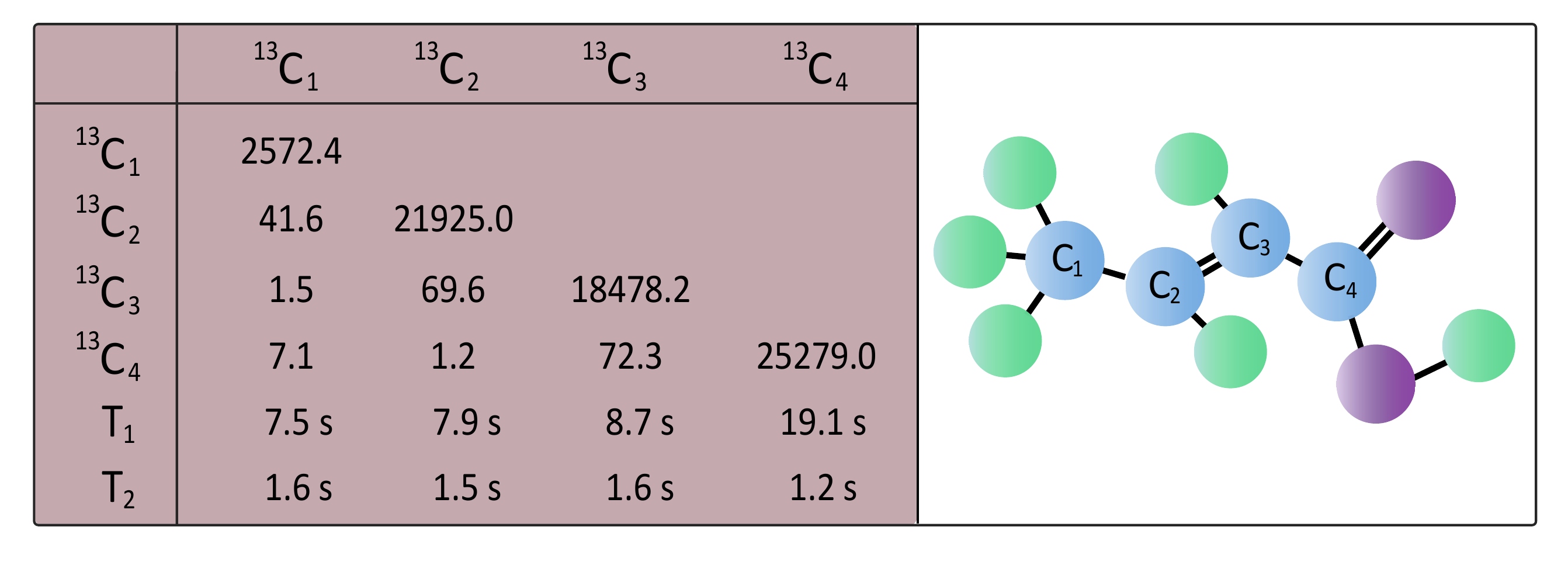}
\caption{Molecular structure and the NMR parameters of the sample. The chemical shifts $\nu_i$ and the scalar coupling strengths $J_{ij}$ are given in the diagonal and off-diagonal segments of the accompanying table, respectively. The longitudinal and transverse relaxation times $T_{1,2}$ (in seconds) are shown at the bottom.}
\label{molecular}
\end{figure}

\textit{Characterization.}--Our experiments were conducted on a Bruker 600 MHz spectrometer ~\cite{Xin_2018,RevModPhys.76.1037,JONES201191,KHANEJA2005296,park2016simulation}, using the 
$^13$C-labeled crotonic acid molecules dissolved in deuterated acetone (d6-acetone) as a quantum simulator.
The four coupled $^{13}$C spins correspond to the four qubits of the quantum simulator. %Throughout the experimental procedures, the methyl group M and H atoms in the molecule were decoupled. 
The internal dynamics of the system %, considered in the rotating frame with a reference frequency $\nu_0$, 
is governed by the internal Hamiltonian
\begin{equation}
\label{eq:NMRHamiltonianSuppl}
\mathcal{H}_{\text{int}} = \pi\sum^4_{i=1}\nu_i\sigma^i_z + \pi\sum^4_{i<j,=1}\frac{J_{ij}}{2}\sigma^i_z\sigma^j_z,
\end{equation}
where $\sigma_z^i$ represents the Pauli-Z matrix of the $i$-th spin, $\nu_i$ is the chemical shift of the $i$th spin and $J_{ij}$ is the J-coupling strength between the $i$-th and $j$-th spins. The specific structure of the four utilized spins, along with their respective parameters, is depicted in Fig.~\ref{molecular}. In this setup, $^{13}$C$_1$ acts as the probe qubit, $^{13}$C$_{2,4}$ constitute the two system qubits (A and C), and $^{13}$C$_3$ is the ancillary qubit to simulate the effect of the channel $\mathcal{M}$.
The system is controlled using a radio-frequency (RF) field, with control Hamiltonian described as 
\begin{equation}
\mathcal{H}_{\text{con}} = - B_1\sum\limits_{i = 1}^4\gamma_i[\cos(\omega_{rf}t+ \phi)\sigma_x^i+\sin(\omega_{rf}t+ \phi)\sigma_y^i],
\end{equation}
where $B_1$, $\omega_{rf}$ and $\phi$ represent the pulse amplitude, frequency and phase of the control pulse, respectively. Single-qubit rotations are achieved by applying RF pulses, while two-qubit gates utilize scalar coupling between different spins combined with the spin echo technique.

\begin{figure}[t]
\centering
\includegraphics[width=0.8\linewidth]{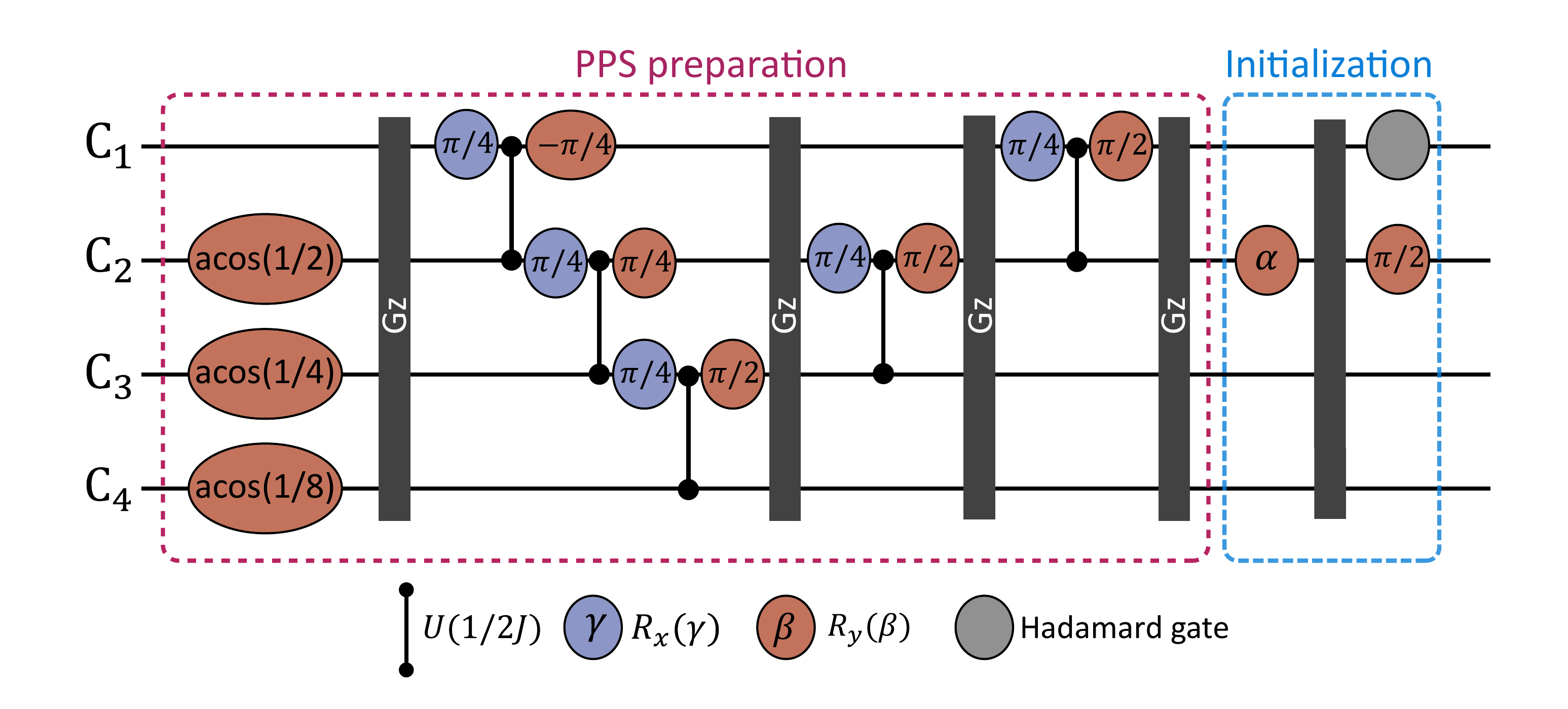}
\caption{The quantum circuit to initialize the four-qubit quantum simulator. The operations encased in the red color dashed box are utilized for generating the pseudo pure state via the spatial averaging method. Following this, the blue dashed box indicates the preparation of the desired initial state of the quantum casual inference experiment.}
\label{circuit}
\end{figure}

\textit{Pseudo-pure state preparation.}--Starting from an initial thermal equilibrium
state, we first create a pseudopure state (PPS).
The thermal equilibrium state of the spin system at room temperature is described by
\begin{equation}
\rho_{\text{eq}} = \frac{1-\epsilon}{16}\mathbb{I} + \epsilon\sum^4_{i=1}\sigma_z^i,
\label{eqstate}
\end{equation}
where $\mathbb{I}$ is the $16 \times 16$ identity matrix, and $\epsilon$, representing polarization, is approximately $10^{-5}$. Several methods exist for system initialization in NMR, including the spatial average method, the selective transition method, the time-spatial method, and the cat-state method. For our experiment, we opted for the spatial average method to initialize the NMR system, employing the pulse sequence depicted in Fig.~\ref{circuit}. In the circuit diagram, the single-qubit rotations are implemented using rf pulses. The two-qubit gates are executed by exploiting the scalar coupling and spin echo technique and optimized utilizing shaped pulses. This pulse sequence transforms the equilibrium state described in Eq.~\eqref{eqstate} into a PPS
\begin{equation}
\rho_{\text{PPS}} = \frac{1-\epsilon'}{16}\mathbb{I} + \epsilon'\op{0000}{0000},
\label{eq:PPS}
\end{equation}
where $\epsilon'$ is the polarization of the PPS. The substantial component of the identity matrix $\mathbb{I}$ remains unchanged under any unitary transformation and is undetectable in NMR experiments; This allows the quantum state to be regarded as the pure state $\op{0000}{0000}$. 

\textit{Initialization.}--In the partial swap channel experiment, we initialize the system to $\rho_0=\ket{+}\bra{+}\otimes\rho_A\otimes\ket{00}\bra{00}$, where $\rho_A=(1-\lambda) \mathbb{I}/2+\lambda\ket{+}\langle+|$. The mixed state $\rho_A$ can be prepared using two single-qubit gates alongside a gradient-field pulse, presented by the circuit depicted in the blue dashed box in Fig.~\ref{circuit}. The first single-qubit rotation, denoted as $R_y(\alpha)$, rotates the qubit to state in the form of $\sqrt{(1+\lambda)/2}\vert0\rangle+\sqrt{(1-\lambda)/2}\vert1\rangle$, where $\alpha = \text{arccos}\sqrt{\langle \sigma_x \rangle^2 + \langle \sigma_y \rangle^2 + \langle \sigma_z \rangle^2}$, 
with $\langle \sigma_i \rangle = \text{Tr}(\rho_A \sigma_i)$. 
Subsequently, a gradient-field pulse is applied to dephase the state, remove the coherence and transform the state into the mixed state, $(1-\lambda)\mathbb{I}/2+\lambda\vert0\rangle\langle0\vert$, achieving the same purity as the target state $\rho_A$. Following this, a second single-qubit rotation $R_y(\pi/2)$ evolves the state into $\rho_A$. Simultaneously, a Hadamard gate is applied to the probe qubit to prepare it in $\vert+\rangle\langle+\vert$.

\subsection{Eigenvalues of Choi Matrices}

Here, we provide additional experimental results for the section `Experiment for $A\to B$, unitary case'. Specifically, we include more experimental data on the eigenvalues of the Choi matrices for the forward and backward processes.

\begin{figure}[t]
\centering
\includegraphics[width=0.8\linewidth]{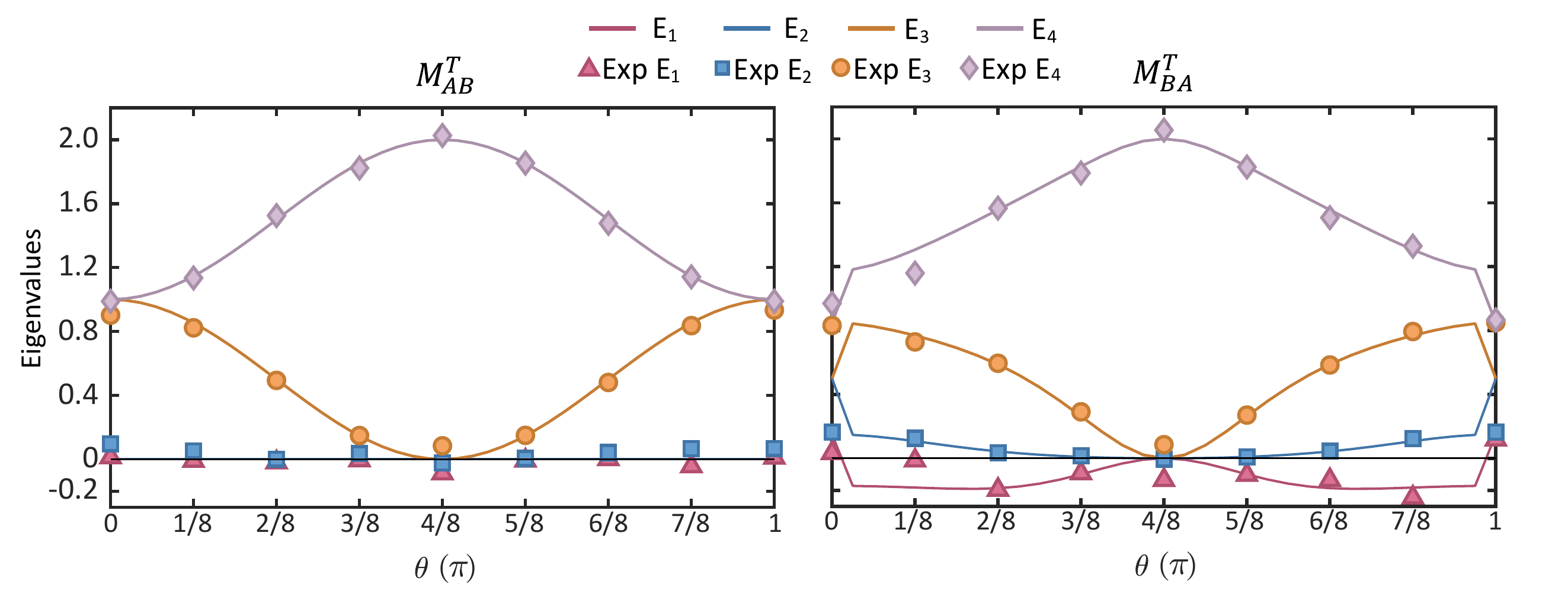}
\caption{The eigenvalues of $M_{AB}^T$ and $M_{BA}^T$ when fixing the input state fixed at $\lambda=0.7$ and the channel strength  $\theta$ varied from $0$ to $\pi$. The solid line represents the theoretical results, while the markers indicate the experimental results.}
\label{SMresult1}
\end{figure}

In our first experiment, we set $\lambda=0.7$ and systematically varied the channel strength $\theta$ over the range from $0$ to $\pi$, measuring the corresponding reduced PDM $R_{AB}$. To validate the effectiveness and stability of the protocol, we calculated the Choi matrices $M_{AB}^{T}$ of the $A\to B$ processes and the corresponding time-reverse Choi matrices $M_{BA}^{T}$ from the experimental PDM $R_{AB}$, allowing us to deduce their eigenvalues. 
The corresponding eigenvalues are shown in  Fig.~\ref{SMresult1}. For $M_{AB}^T$, as the channel strength $\theta$ varies, its eigenvalues remain positive. In contrast, for $M_{BA}^T$, negative eigenvalues are shown. 
These results indicate that the data $R_{AB}$ is compatible with a cause-effect mechanism with $A$ being the cause and $B$ being the effect, which is consistent with our model.

\begin{figure}[t]
\centering
\includegraphics[width=0.8\linewidth]{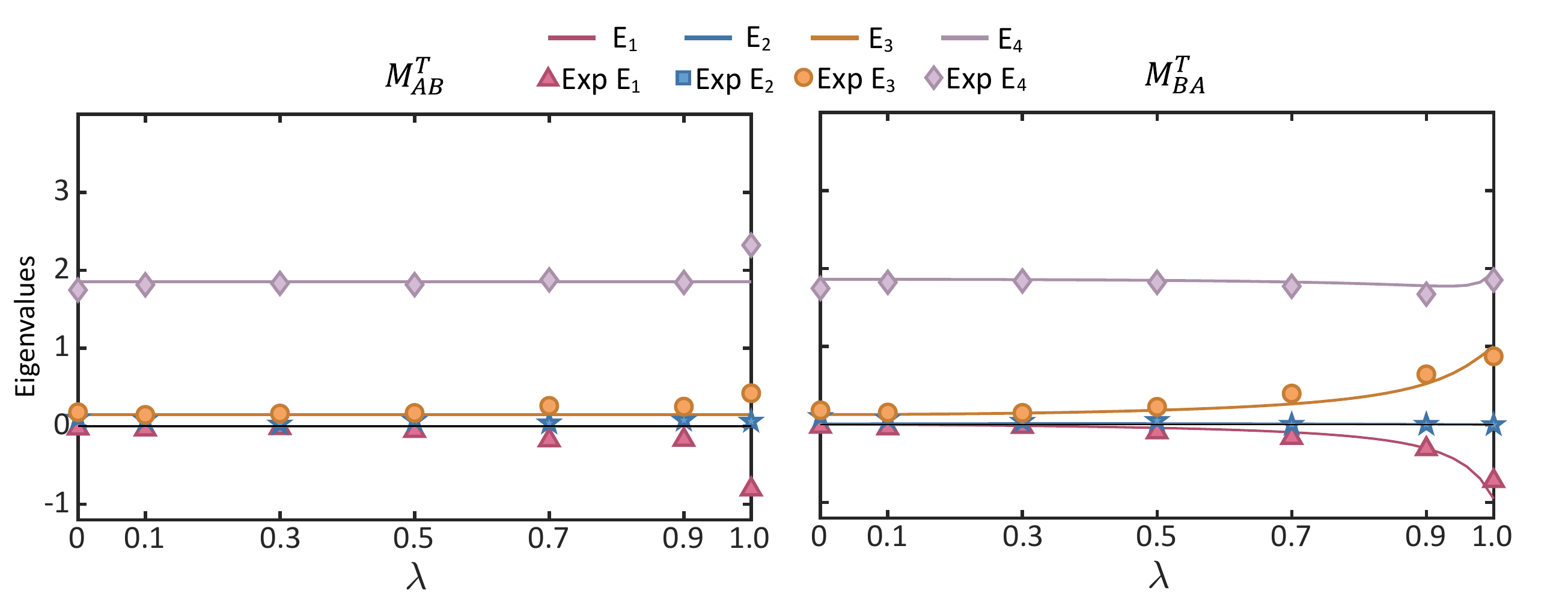}
\caption{The eigenvalues of $M_{AB}^T$ and $M_{BA}^T$ when fixing the channel strength at $\theta=3\pi/8$ and varying the polarization  of initial state $\lambda$ from $0$ to $1$. The solid line represents the theoretical results, while the markers indicate the experimental results.}
\label{SMresult2}
\end{figure}

In our second experiment, we varied the polarization parameter $\lambda$ while fixing the strength of the channel at $\theta=3\pi/8$ and measured the corresponding reduced PDM $R_{AB}$. We also calculated the eigenvalues of the Choi matrices  $M_{AB}^T$ and $M_{BA}^T$, with results displayed in Fig.~\ref{SMresult2}. When the initial state is varied, different initial states pass through the same channel. Consequently, the corresponding matrix $M_{AB}^T$ computed using PDM $R_{AB}$ remains invariant, and the eigenvalues of $M_{AB}^T$ are constant and positive. However, for its time-reversed matrix
$M_{BA}^T$, different $M_{BA}^T$ are obtained. The magnitude of the negative eigenvalues of $M_{BA}^T$ increases with increasing polarization $\lambda$.

\section{Experimental Details for `Experiment for $A \to B$, Fully Decohering Channel'}

The setup for this experiment is similar to the experiment in the section `Experiment for $A \to B$, unitary case'. Interested readers please refer to Section.~\ref{metho}. In the following, we provide details about the fully decohering channel and its realization.

In this experiment, the functional relation from $A$ to $B$ is described by a fully decohering channel. To be concrete, a quantum state $\rho$ at $A$ passing through a fully decohering channel arrives at $B$. The fully decohering channel is given by
\begin{equation}
    \mathcal{L}_{A\to B}(\rho)=\langle0|\rho_A\ket{0}\ket{0}_B\langle0|+\langle1|\rho_A\ket{1}\ket{1}_B\langle1|. 
    \label{FDC}
\end{equation}
%Due to the limited number of qubits of the quantum simulator in our experiments, we could not use the scattering circuit to measure the PDM. 

\begin{figure}[t]
\centering
\includegraphics[width=0.6\linewidth]{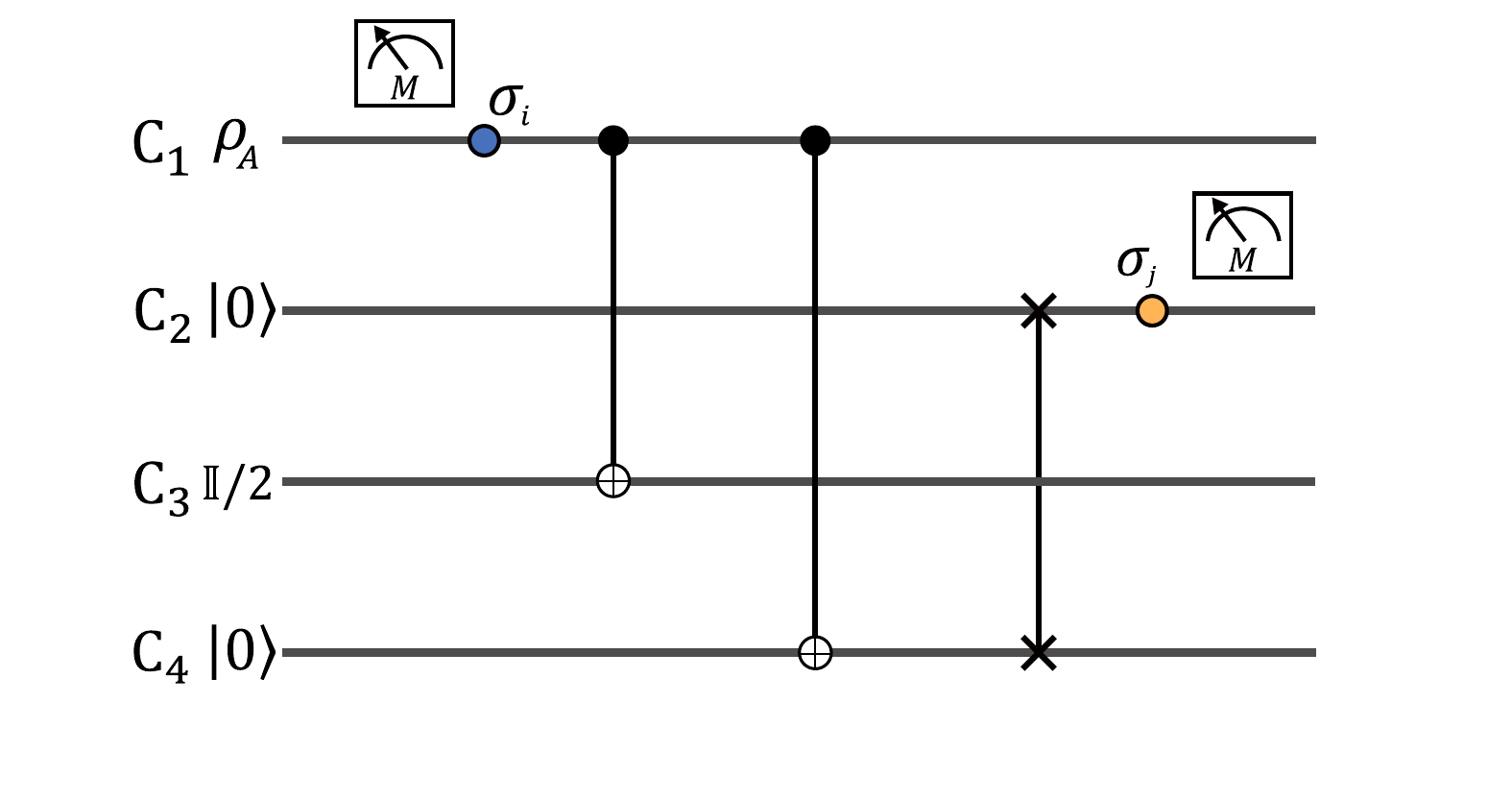}
\caption{Quantum circuit for measuring the two-time correlators $\langle \sigma_i^A, \sigma_j^B \rangle$ in the cause-effect mechanism with the fully decohering channel.}
\label{SMdpcircuit}
\end{figure}

The circuit, consisting of four qubits to realize the decohering channel, is shown in Fig.~\ref{SMdpcircuit}. 
The first spin, C$_1$, carries the input state of the channel $\rho_A$, while the final state of the channel is output by the second qubit, C$_2$. The other two qubits, C$_3$ and C$_4$, are used to simulate the fully decoherent channel.
The initial state of the system is $\rho_{A}\otimes\vert0\rangle\langle0\vert\otimes\frac{\mathbb{I}}{2}\otimes\vert0\rangle\langle0\vert$, where $\rho_A=(1-\lambda)\frac{\mathbb{I}}{2}+\lambda\ket{+}\langle+|$.
This circuit also contains three quantum gates. 
The first gate is a CNOT gate, with C$_1$ as the control qubit and C$_3$ as the target qubit. The reduced density matrix of C$_1$ after that is
\begin{equation}\rho^d_{A}=\inp{0|\rho_A|0}\op{0}{0}+\inp{1|\rho_A|1}\op{1}{1}.\end{equation} 
The second gate is also a CNOT gate with C$_1$ as the control qubit, while the target qubit is C$_4$, resulting in a final state 
%Subsequently, $\rho^d_{A}$ serves again as the control qubit to perform CNOT operation with the $\ket{0}$ state, resulting in the final state on quantum qubits C$_1$ and C$_4$ being
\begin{equation}\rho_{1,4}=\inp{0|\rho_A|0}\op{00}{00}+\inp{1|\rho_A|1}\op{11}{11}.\end{equation}
This operation essentially copies $\rho^d_{A}$ onto the C$_4$. 
Finally, a swap operation between C$_2$ and C$_4$ yields \begin{equation}\rho^t_{AB}=\langle0|\rho_A\ket{0}\ket{0}_A\langle0|\otimes\ket{0}_B\langle0|+\langle1|\rho_A\ket{1}_A\ket{1}_A\langle1|\otimes\ket{1}_B\langle1|.\end{equation} 
When tracing out the first qubit, the reduced state of C$_2$ turns into 
Eq.~\eqref{FDC}, as desired.% which suggests a successful realization of the fully decohering channel.

Due to the limited number of qubits of the quantum simulator in our experiments, we could not use the scattering circuit to measure the PDM. 
In the experiment, we first prepared spin C$_1$ in the state $\rho_A$, then implemented the coarse-grained Pauli measurement before and after the channel. These data are then used to calculate the expectation values $\langle \sigma_i^A \sigma_j^B \rangle$, therefore constructing the PDM $R_{AB}$. Of course, the PDM can also be measured using a scattering circuit on a quantum simulator with a higher number of qubits.

\begin{figure}[t]
\centering
\includegraphics[width=0.8\linewidth]{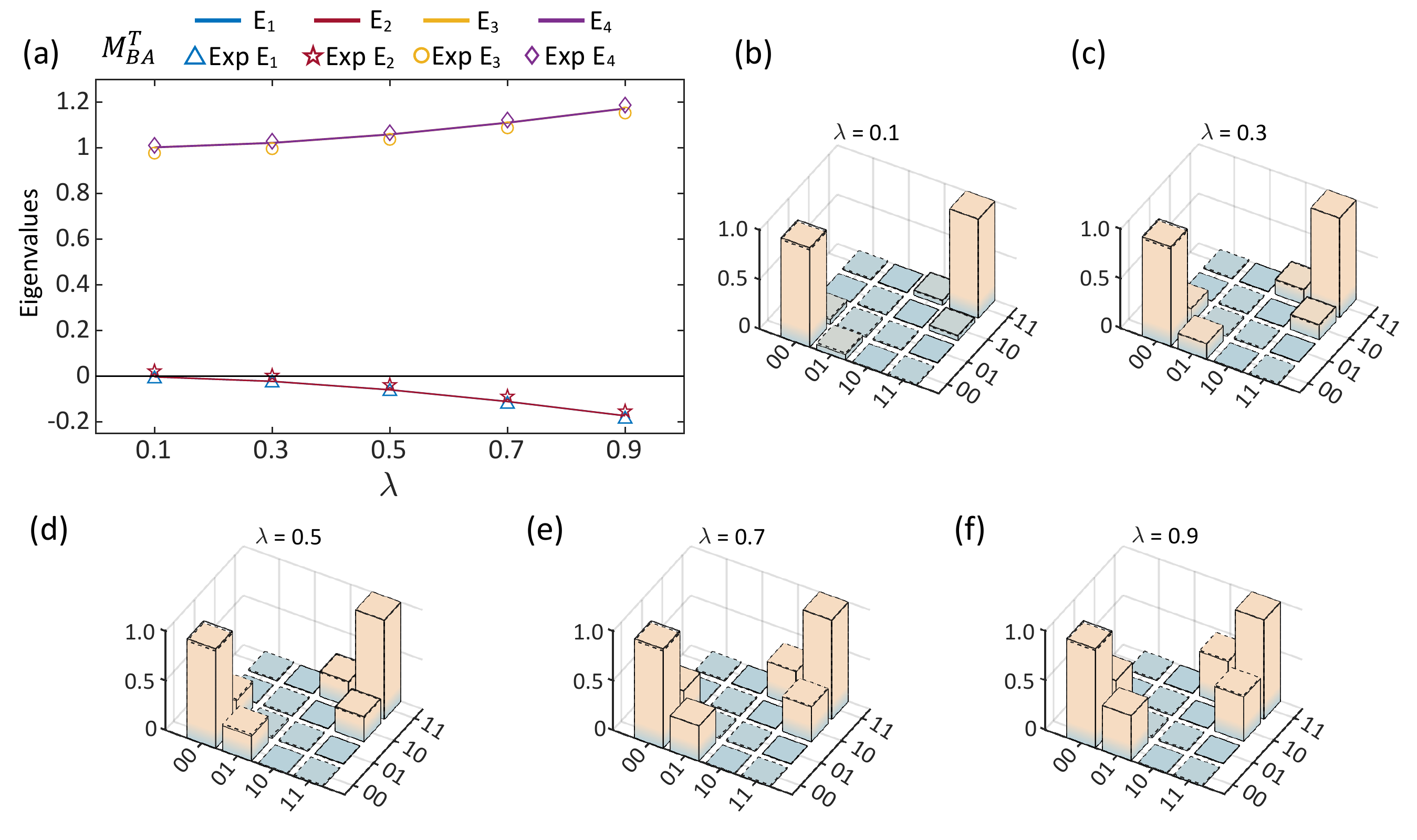}
\caption{(a) The eigenvalues of the backward Choi matrix $M_{BA}^{T_B}$ vary with $\lambda$. Markers indicate eigenvalue points obtained from the experiment, while the solid line depicts the corresponding theoretical results. (b-f) The Choi matrices $M^{T_B}_{BA}$ corresponding to different $\lambda$. The black solid line represents the theoretical results, while the color bars with dashed lines represent the experimental results.}
\label{SMresult3}
\end{figure}

To determine the cause and effect, we study the Choi matrix of the forward and backward processes. In the experiment, we fixed the fully decoherent channel but varied the input states, i.e., varying $\lambda$ from $0$ to $1$. The forward Choi matrix $M_{AB}^T$ remains the same for different values of $\lambda$, while the time-reversed Choi matrices $M_{BA}^T$ change. The eigenvalues for $M_{BA}^T$ and $M_{BA}^T$ corresponding to different $\lambda$ are illustrated in Fig.~\ref{SMresult3}. The magnitude of the negative eigenvalues of $M_{BA}^T$ increases with increasing polarization $\lambda$. This indicates that the temporal direction is  $A\to B$.

\section{Experiments for the Common-Cause Mechanism}

\begin{figure}[htbp]
\centering
\includegraphics[width=0.8\linewidth]{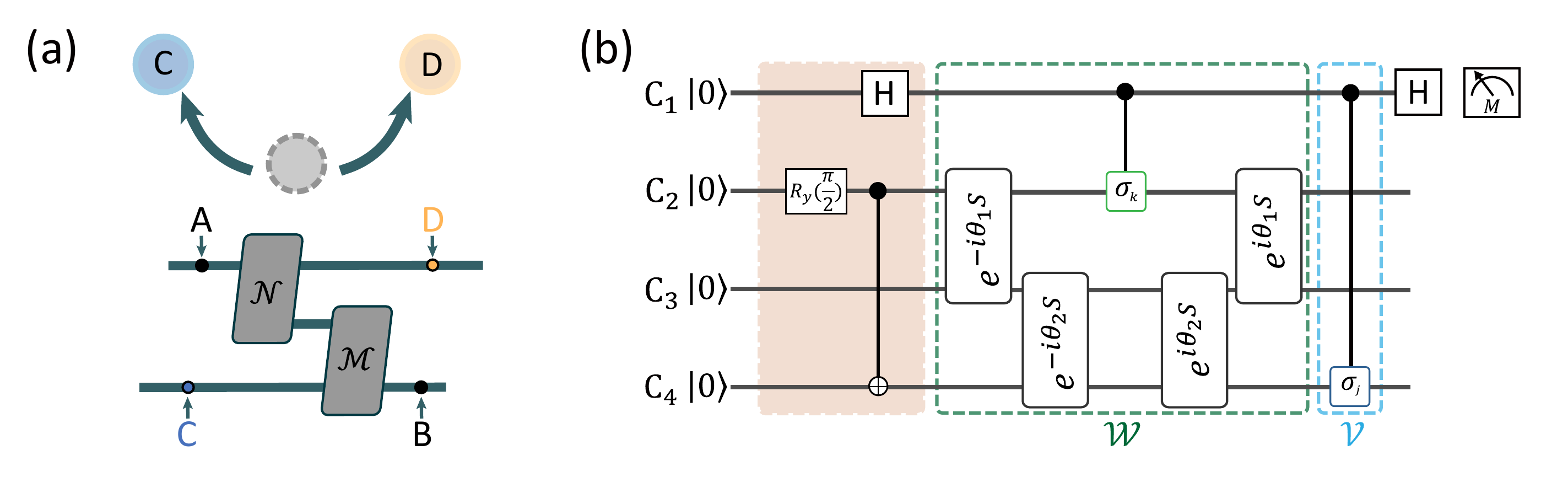}
\caption{(a) The common-cause structure between $C$ and $D$ and the corresponding sampling locations in the circuit where the channels $\mathcal{N} = e^{-i\theta_{1} S}$ and $\mathcal{M} = e^{-i\theta_{2} S}$. (b) Experimental quantum scattering ciccuit. The circuit is designed to measure expectation values of the form $\langle \sigma_j^C, \sigma_k^{D} \rangle$, which are used to reconstruct the PDM $R_{CD}$.}
\label{fig:circuit c1d2 with ComCau}
\end{figure}

We conducted an experiment targeting the common-cause mechanism. The corresponding causal structure and experimental circuit are shown in Fig.~\ref{fig:circuit c1d2 with ComCau}. The structure of the experiment circuit is similar to the scheme described in Fig.2 of the main text, but with a different initial state, different interactions and different measurement locations.
The initial state is $\ket{+}\otimes[\frac{1}{\sqrt{2}}\ket{000}+\frac{1}{\sqrt{2}}\ket{101}]$, and the channels are $\mathcal{N} = e^{-i\theta_{1} S}$ and $\mathcal{M} = e^{-i\theta_{2} S}$, with $\theta_1$ and $\theta_2$ being  the parameters.

In this experiment, the circuit is modified to extract the expectation values $\langle \sigma_j^C, \sigma_k^D \rangle$, which are then used to reconstruct the PDM $R_{CD}$ for causal inference.
As depicted in Fig.~\ref{fig:circuit c1d2 with ComCau}(a), system $C$ does not exert any direct causal influence on system $D$ after the channel operations—any observed correlation between them originates solely from the shared common cause.

\begin{figure}[htbp]
\centering
\includegraphics[width=0.8\linewidth]{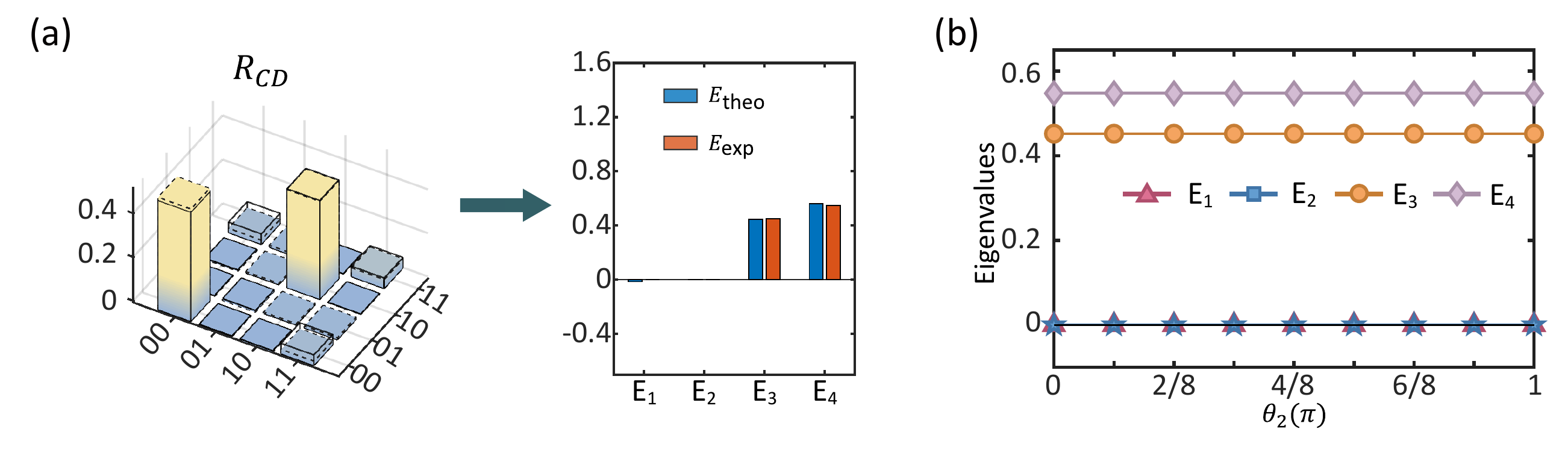}
\caption{(a) The experimental result of PDM $R_{CD}$ and its corresponding eigenvalues. (b) The eigenvalues of $R_{CD}$ as a function of $\theta_2$, with $\theta_1 =2\pi/5$.}
\label{fig:results c1d2 with ComCau}
\end{figure}

The experimental results are presented in Fig.~\ref{fig:results c1d2 with ComCau}(a). We find that the negativity of PDM $R_{CD}$  $f(R_{CD})=0$, indicating that $R_{CD}$ is positive semidefinite.
Furthermore, we varied the channel parameter $\theta_2$ of $\mathcal{M}$ from 0 to $\pi$, and the simulation results show that the eigenvalues of $R_{CD}$ remain invariant.  Therefore, we assign the common-cause mechanism to $C$ and $D$.  The observed correlations can therefore be attributed solely to the common cause originating from the entangled initial state.

%%%%%%%%%%%%%%%%%%%%%%
\section{Experiments for the Mixture of the Common-Cause and Cause-Effect Mechanisms}
\begin{figure}[htbp]
\centering
\includegraphics[width=0.8\linewidth]{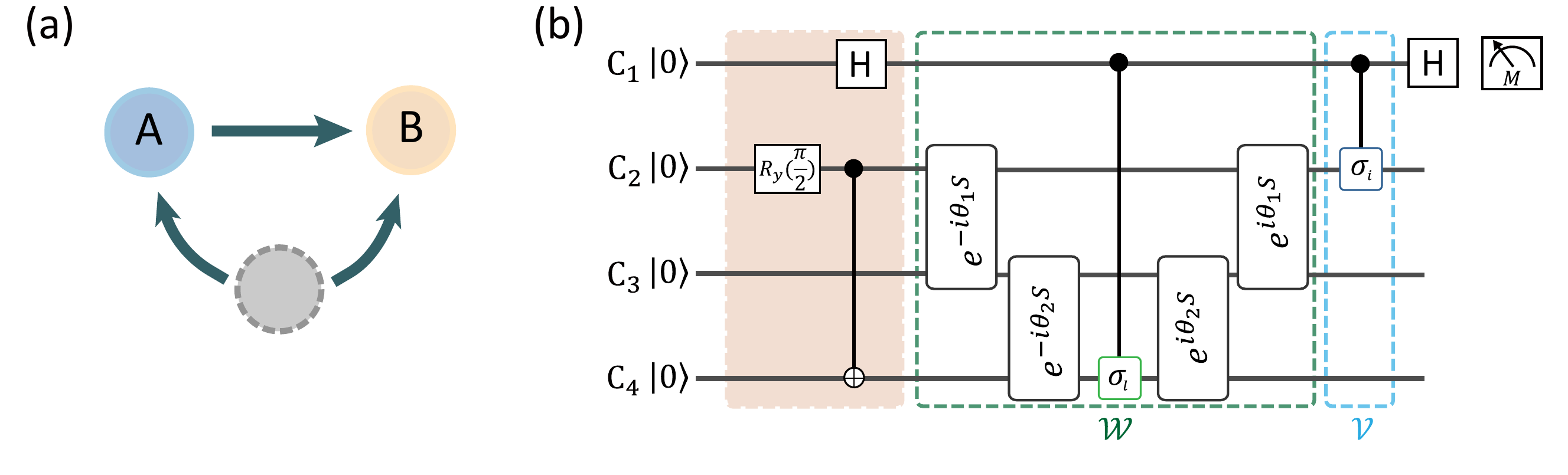}
\caption{(a) Causal structure illustrating a direct influence from system A to system B, along with a shared common cause (gray dashed node) affecting both. Nodes represent quantum systems, and directed edges represent causal influence. (b) Experimental quantum circuit on the four spins. The aim of the circuit is to measure expectation values $\langle \sigma_i^A,\sigma_l^{B}\rangle$.}
\label{fig:circuit a1b2 with ComCau}
\end{figure}

We conducted the experiment targeting the mixture of the common-cause mechanism and the cause-effect mechanisms. The causal structure between $A$ and $B$ is illustrated in Fig.~\ref{fig:circuit a1b2 with ComCau}~(a). The circuit is similar to Fig.2~(a) of the main text, but initial correlation between $A$ and $C$ are presented as a common cause.

The corresponding experimental circuit is shown in Fig.~\ref{fig:circuit a1b2 with ComCau}~(b). After the state preparation steps indicated in the yellow box, the initial state of the four-qubit quantum processor is given by $\ket{+}\otimes[\frac{1}{\sqrt{2}}\ket{000}+\frac{1}{\sqrt{2}}\ket{101}]$, i.e., $AC$ is the Bell state.
As shown in Fig~2(a), we set  $\mathcal{N}=e^{-i\theta_{1} S} $ and $\mathcal{M}=e^{-i\theta_{2} S}$ .
We then employ the scattering circuit to obtain two-time correlators $\langle \sigma_i^A,\sigma_l^{B}\rangle$, which are used to construct PDM $R_{AB}$.

\begin{figure}[t]
\centering
\includegraphics[width=0.8\linewidth]{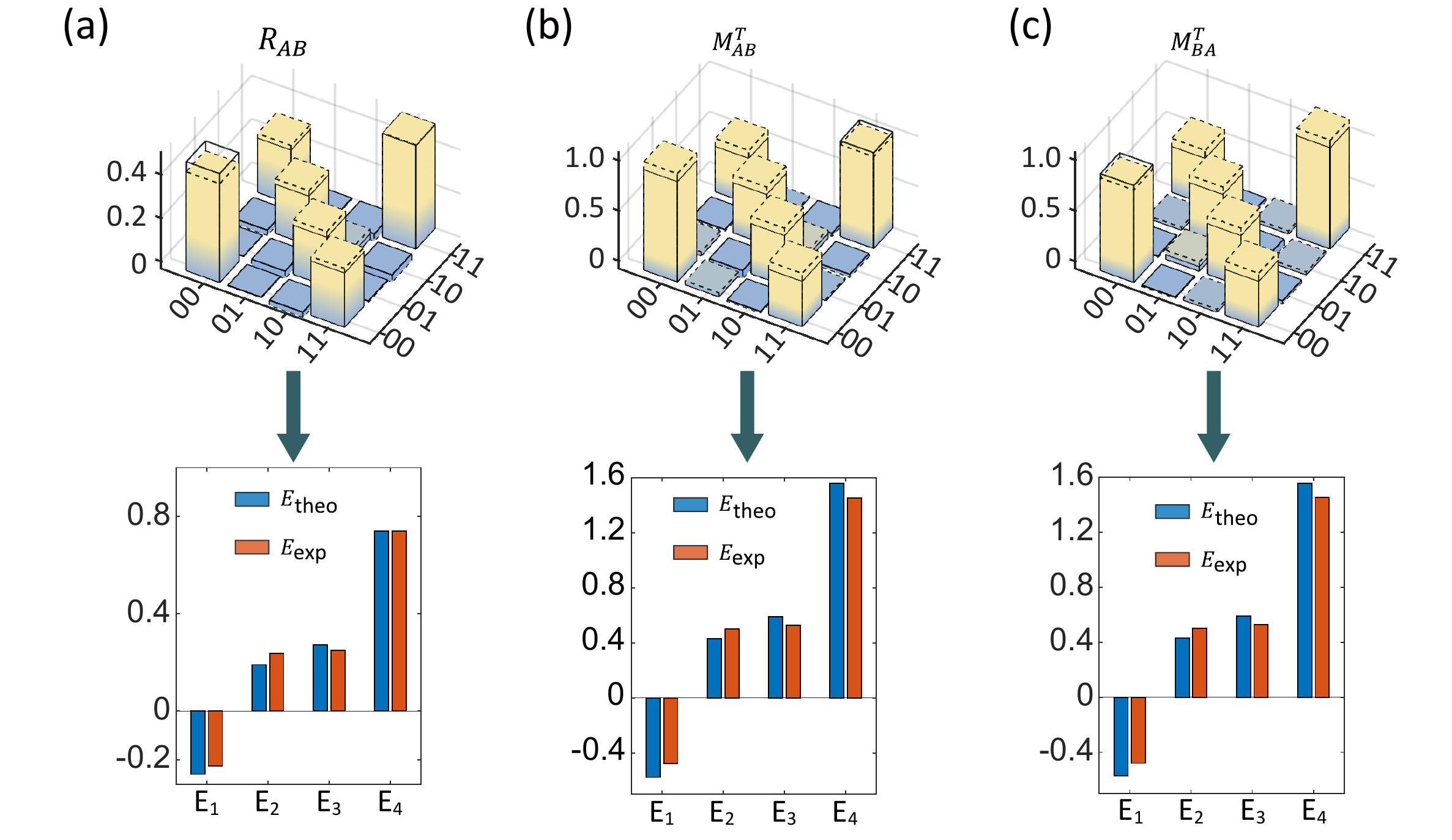}
\caption{Results for $\theta_1= 2\pi/5$ and $\theta_2=\pi/4$. The top row shows the constructed matrices: (a) $R_{AB}$, (b) $M_{AB}^T$ and (c) $M_{BA}^T$. In each case, the black solid line represents the theoretical results, while the color bars with dashed lines represent the experimental results. The bottom row presents the corresponding eigenvalues of each matrix. Blue bars represent theoretical values ($E_{\text{theo}}$), and red bars represent experimental results ($E_{\text{exp}}$). Note that $M^{T_B}_{AB}$ and $M^{T_B}_{BA}$ are almost identical due to the symmetry between the forward and backward processes.}
\label{fig:result a1b2 with ComCau}
\end{figure}

Setting $\theta_1= 2\pi/5$ and $\theta_2= \pi/4$, the experimental results for PDM $R_{AB}$, the Choi matrices $M_{AB}^T$ and $M_{BA}^T$ along with their corresponding eigenvalues, are shown in Figs.~\ref{fig:result a1b2 with ComCau}(a–c). According to our quantum causal inference protocol, we first evaluate the negativity $f(\cdot)$ of $R_{AB}$ and find out that $f(R_{AB})>0$, therefore there must be causal influence between $A$ and $B$. Then, we calculate the Choi matrices for the forward and backward processes and find out that $f(M^{T_B}_{AB})>0$ and $f(M^{T_B}_{BA})>0$, implying the presence of a common cause. Thus we infer the causal structure between $A$ and $B$ is the mixture of the two mechanisms.

\section{Experimental Error in $R_{AB}$}

\begin{figure}[t]
\centering
\includegraphics[width=1\linewidth]{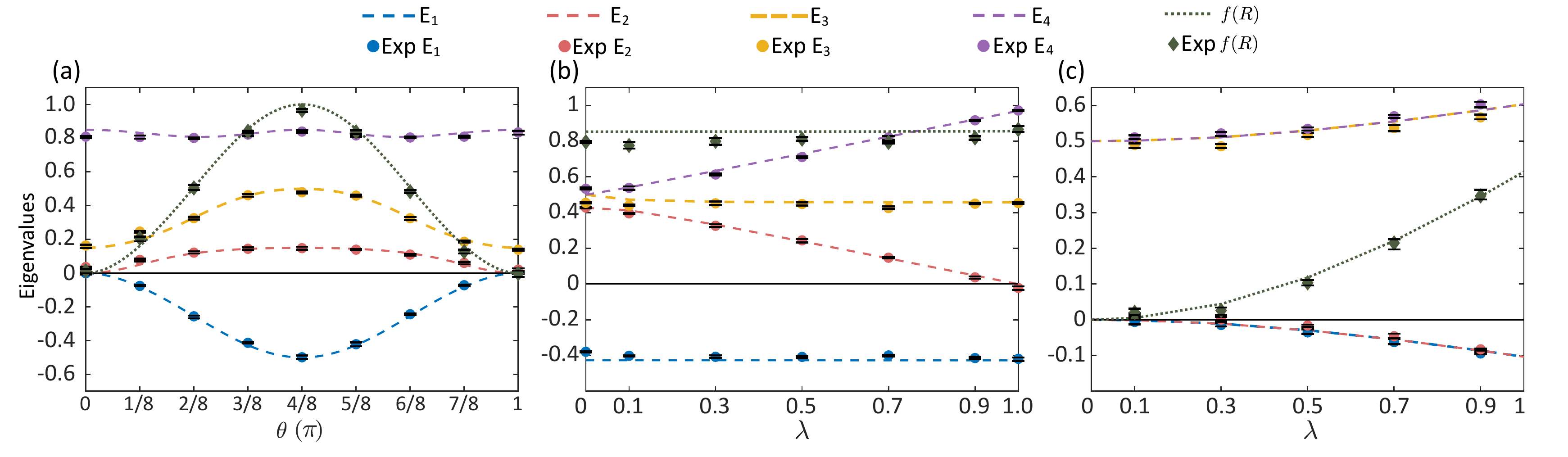}
\caption{The eigenvalues E$_i$ and $f(R_{AB})$ of $R_{AB}$ in the main text with error bars derived from five experimental repetitions. (a-b) The $R_{AB}$ for different parameters obtained from partial swap channel. (c) The $R_{AB}$ for different $\lambda$ obtained from fully decohering channel.}
\label{Error}
\end{figure}

In NMR platforms, the experimental samples do not consist of a single molecule, but rather a system composed of a thermodynamic limit number of identical molecules.
Consequently, the quantum state measurements performed in NMR are measurements of ensemble averages, which inherently stabilize against considerable fluctuations. Moreover, our experimental duration is controlled to remain within the decoherence time. Thus the effects of decoherence are negligible, leading to highly reliable results. The experimental error primarily arises from imperfections in quantum control.
We have incorporated error bars in all $R_{AB}$ mentioned in the main text, derived from five experimental repetitions, as illustrated in Fig.~\ref{Error}. Each experimental result is almost constant, resulting in error bars that fall within the size of the marker.

\section{Extracting $M$ when the Initial State is Rank-Deficient}

When the initial state of the PDM, $ R_{AB}=(\rho_A \otimes \mathbb{I}_B) M_{AB} +M_{AB}  (\rho_A \otimes \mathbb{I}_B)$, is full-rank, we have provided a method in the main text to extract information about the dynamics. Here, we present an algorithm designed to extract the dynamics from the PDM when the initial state is rank-deficient, aiming to obtain the corresponding Choi matrix 
$M^{T_B}_{AB} $ that is the least negative.

As found in Refs.~\cite{liu2023inferring,liu2023quantum}, $M_{AB}$ is not unique for the case of $\rho_A$ being rank deficient.  The following semidefinite programming (SDP) is proposed to give a CJ matrix $M_{AB}$ with its Choi matrix $M^{T_B}_{AB}$ being the least negative,
\begin{equation}
\begin{aligned}
    & \text{minimize } \tr M_{-}^{T_{A}}
    \\
    & \text{subject to }
    \left\{\begin{aligned}
     & M = M^{\dagger}, \\
     & R = (\rho\otimes \mathbb{I}) M +M (\rho\otimes \mathbb{I}), \\
     & M^{T_{B}}=M^{T_{B}}_{+}-M^{T_{B}}_{-}, \\
     & \tr_{B} M = \mathbb{I}_{B}, \\
     & M^{T_{A}}_{+} \geq 0 , \quad M^{T_{B}}_{-} \geq 0.
    \end{aligned} \right.
\end{aligned}
\end{equation}
This SDP can be solved numerically using standard convex-optimization modeling tools together with a suitable SDP solver that supports partial transpose. In our implementation, we formulate the SDP in MATLAB and solve it using the {\tt CVX} package (for convex optimization) together with {\tt QETLAB} (for the partial transpose). There are various options. Here, we establish the SDP in Matlab and apply the Packages {\tt{CVX}} (for convex optimization) and {\tt{QETLAB}} (for implementing partial transpose).
The code for the SDP problem is then provided by the function {\tt{causalinfer}} as follows:
\begin{lstlisting}
function N = causalinfer(R,d)
rho=kron(PartialTrace(R,2,[d,d]),eye(d));
cvx_begin sdp
variable Np(d*d,d*d) hermitian
variable Nn(d*d,d*d) hermitian
N=Np-Nn;
Nptp=PartialTranspose(Np,1,[d,d]);
Nntp=PartialTranspose(Nn,1,[d,d]);
minimize( trace(Nntp) )
0.5*(rho*N+N*rho)==R;
PartialTrace(N,2,[d,d])==eye(d);
Nptp >= 0;
Nntp >= 0;
cvx_end
N=N+zeros(d*d);
end
\end{lstlisting}
Then the optimal value of SDP and the corresponding solution are given by typing {\tt{causalinfer(R,d)}}.

\end{document}